\documentclass{elsarticle}
\usepackage{color}
\usepackage{graphicx}
\usepackage{booktabs}
\usepackage{pifont}
\usepackage{amsthm,mathtools}
\usepackage{longtable,ltcaption}
\usepackage{rotating}
\usepackage{multicol}
\usepackage{multirow}
\usepackage{lipsum}
\usepackage{bbm}
\usepackage{amsmath}
\usepackage{amssymb}
\usepackage{xcolor}

\def\reals{\mathbbm{R}}
\def\B{{\cal B}}
\def\F{{\cal F}}
\def\L{{\cal L}}
\def\S{{\cal S}}

\newtheorem{theorem}{Theorem}

\definecolor{blue}{RGB}{1,1,1}

\begin{document}

\begin{frontmatter}

\title{Stochastic modelling of football matches}

\author[1]{Luiz Fernando G. N. Maia}
\author[2]{Teemu Pennanen}
\author[1]{Moacyr A.H.B. da Silva}
\author[1]{Rodrigo S. Targino}

\address[1]{School of Applied Mathematics (EMAp), Getulio Vargas Foundation (FGV)}
\address[2]{Department of Mathematics, King’s College London}

\begin{abstract}
This paper develops a general framework for stochastic modeling of goals and other events in football (soccer) matches. The events are modelled as Cox processes (doubly stochastic Poisson processes) where the event intensities may depend on all the modeled events as well as external factors. The model has a strictly concave log-likelihood function which facilitates its fitting to observed data. Besides event times, the model describes the random lengths of stoppage times which can have a strong influence on the final score of a match. The model is illustrated on eight years of data from Campeonato Brasileiro de Futebol Série A. We find that dynamic regressors significantly improve the in-game predictive power of the model. In particular, a) when a team receives a red card, its goal intensity decreases more than 30\%; b) the goal rate of a team increases by 10\% if it is losing by one goal and by 20\% if its losing by two goals; and c) when the goal difference at the end of the second half is less than or equal to one, the stoppage time is on average more than one minute longer than in matches with a difference of two goals.
\end{abstract}

\begin{keyword} Stochastic football modelling \sep In-game forecasts \sep Cox Process \sep Convex optimization
\end{keyword}

\end{frontmatter}

\section{Introduction}
In recent years, the field of sports analytics has gained 
immense traction, with a particular focus on predicting the
outcomes of football matches. The ability to forecast match 
results accurately holds significant implications for 
various stakeholders, including sports enthusiasts,
betting markets, and team management. Being
a dynamic and unpredictable sport, football presents unique 
challenges in developing robust forecasting models. 

Several approaches have been proposed for assigning probabilities
to different outcomes of a football match. Arguably the most influential one is that of Maher~\cite{maher1982modelling} where 
the numbers of goals scored by each team are modelled as independent Poisson variables. Several modifications have been proposed in order to relax the independence assumption which is not supported by data; see e.g.\
\cite{dixon1997modelling, analysis_karlis_2003,modelling_mchale_2011, bivariate_boshnakov_2017}. Despite allowing for dependence, the models in the above references do not explain the source of the dependence.

Dixon and Robinson~\cite{dixon1998birth} introduced a dynamic model where the timing of goals during a game is modelled as a doubly stochastic Poisson process where the goal intensities may depend on the state of the game. Such a model induces a natural dependency structure for the final scores of the two teams. A closely related approach was taken by Titman et al.~\cite{tcrg} who used a Weibull proportional hazards formulation for goals as well as bookings (red and yellow cards). They found that both the current score as well as the number of red cards at a particular time during a game had a significant influence on the goal intensities. The significance of red cards on the match score has been studied also in \cite{ten_ridder_1994, consequences_bareli_2006, 
estimating_vecer_2009}.

The present paper proposes a general class of dynamic stochastic 
models to describe relevant events during a football match. Besides 
goals, the modelled events could include red and yellow 
cards, injuries and anything else that is believed to affect the 
final score of a match. The event intensities may depend on dynamic regressors, such as the current score, number of red cards awarded, injuries etc. Allowing for time inhomogenious regressors, our model can be viewed as an extension of both the doubly stochastic Poisson model of \cite{dixon1998birth} as well as the Weibull proportional hazards model of \cite{tcrg}.

Besides event times,
our model describes the distribution of the stoppage times that the 
referee may award at the end of each half of a match. In Campeonato 
Brasileiro de Futebol Série A, for example, more than 7\% of the 
total duration of games have consisted of stoppage times. The 
significance of stoppage times to the final score of a game is 
even higher than that since the goal intensities tend to be at 
their highest towards the end of a game. 

An important feature of our approach is that 
it results in log-likelihood functions that are strictly concave 
in their parameters; see Theorem \ref{thm:concave}, below. This greatly facilitates the fitting of our models to historical data as it guarantees unique parameter estimates and allows for the use of numerical techniques of convex optimization in model fitting. To fit a model with 78 parameters to data from 3\,039 matches it takes less than 10 seconds in a Intel(R) Core i7-10700F CPU @ 2.90GHz (16 CPUs) and 32GB of RAM; see Section~\ref{sec:fitting} below. 

We illustrate the modelling approach by fitting different models to eight seasons of data from the Campeonato Brasileiro de Futebol Série A. We find that dynamic regressors improve the in-game predictive power of the model significantly. In particular, 
\begin{enumerate}
    \item 
    when a team receives a red card, its goal intensity decreases more than 30\%, 
    \item 
    the goal rate of a team increases by 10\% if it is losing by one goal and by 20\% if its losing by two goals,
    \item
    when the goal difference at the end of the second half is less than or equal to one, the stoppage time is on average more than one minute longer than in matches with a difference of two goals.
\end{enumerate}
In out-of-sample test, the best model predicts that 27.48\% of the matches would end as draws which is very close to the observed proportion of 27.42\%.

The rest of the paper is organized as follows. Section~\ref{sec:data} describes a data set that will be analyzed later in the paper. 
Sections~\ref{sec:notation} and \ref{sec:params} describe the general modelling approach and summarizes its main properties. 
Section~\ref{sec:params} presents the particular parametrizations that will be used in the numerical studies later in the paper. Section~\ref{sec:fitting} gives the results on parameter estimation and Section~\ref{sec:forecast} studies model performance in forecasting.

\section{The data}\label{sec:data}

The data used in this paper was collected from three sources and corresponds to the seasons 2015 to 2022 of the Campeonato Brasileiro de Futebol Série A. Information about the matches, goals and red cards was collected from the Flashscore\footnote{https://www.flashscore.com/} website. Stoppage time data was collected from the official match reports present in the Confederação Brasileira de Futebol\footnote{https://www.cbf.com.br/} website. Finally, market values of the players that started each match was obtained from Transfermarkt\footnote{https://www.transfermarkt.com/}.

In the data set, we have a total of 3\,039 matches, 7\,126 goals, 4\,248 of which were from the home team and 717 red cards, 434 of which were received by players of the away team. Red cards awarded to coaches and players on the bench were ignored.

\begin{figure} [ht] 
\caption{Scores' histogram}
\label{scores} 
\centering
\includegraphics[width=0.95\textwidth]{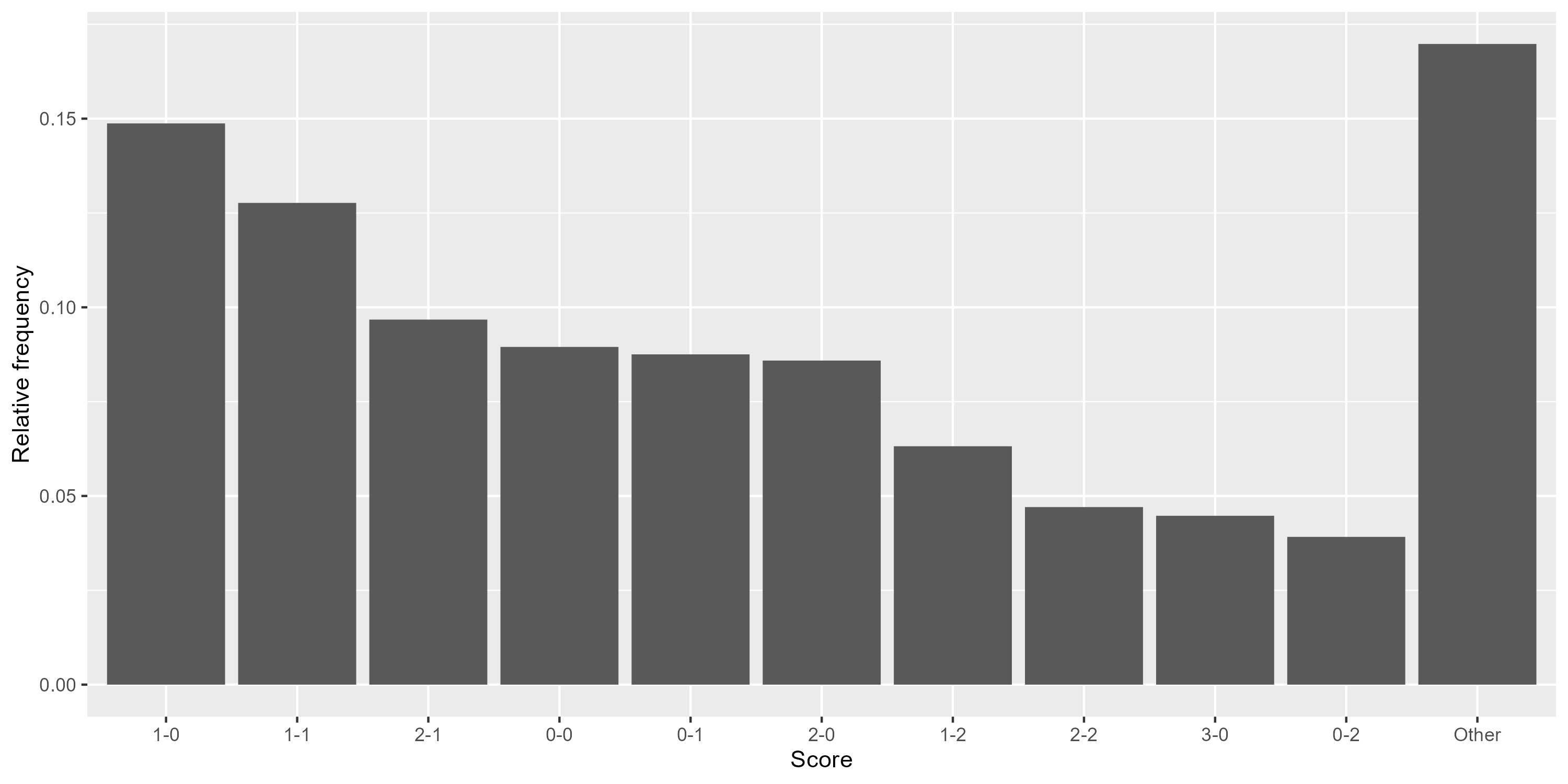}
\end{figure}

\begin{figure} [ht] 
\caption{Frequency of goals scored by home and away teams}
\label{marginal scores} 
\centering
\includegraphics[width=0.95\textwidth]{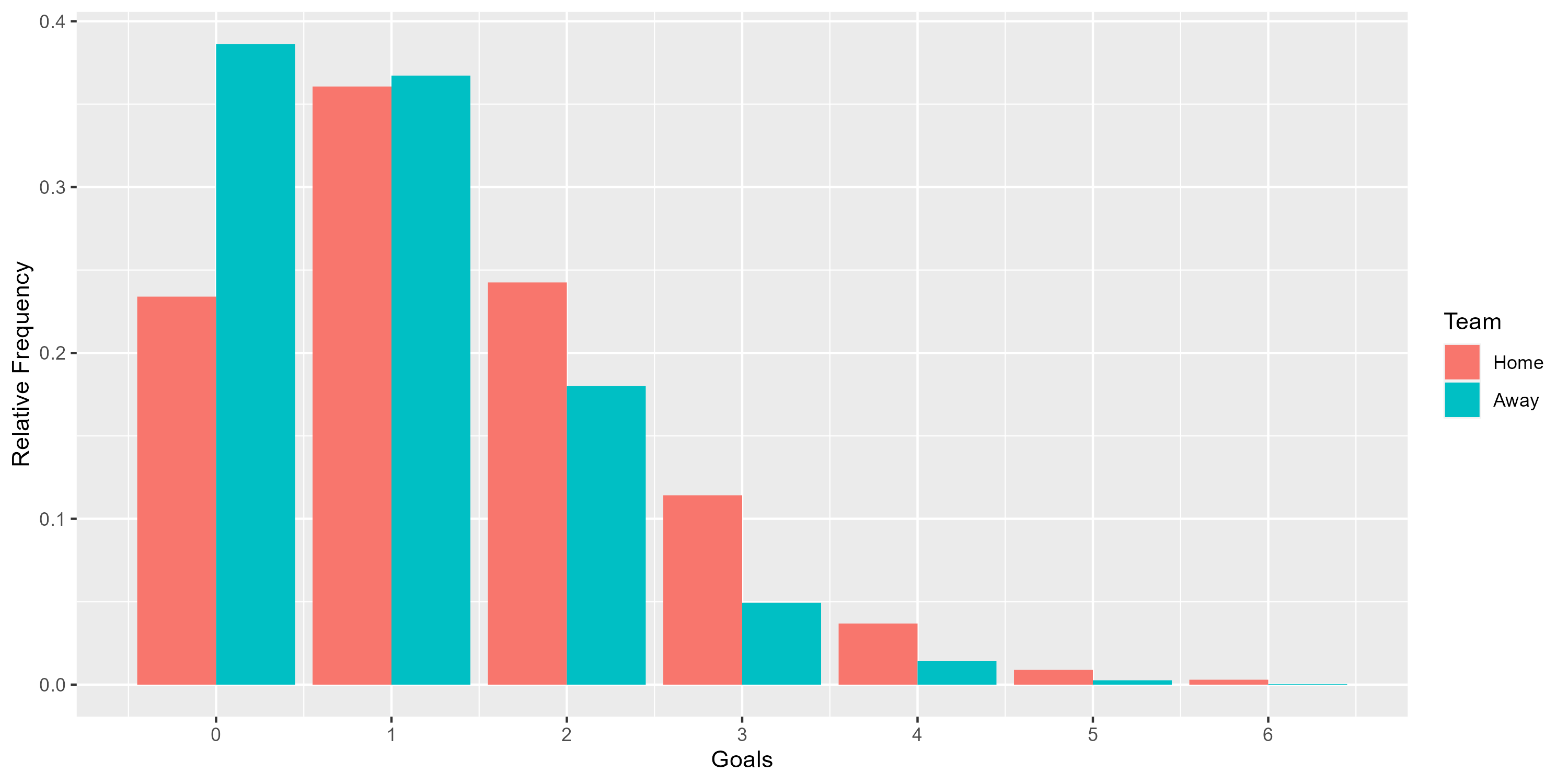}
\end{figure}

Considering all matches in the data set, the home team won 1\,468 (48.3\%) matches, the away team won 746 (24.5\%) matches and 825 (27.1\%) matches ended up in a draw. The 10 most common scores are shown in Figure \ref{scores}. A win 1 to 0 for the home team is the most common score, happening in 14.9\% of the matches. However, there's some variety in the scores since the ``Other'' category has 17.0\% of relative frequency. Figure \ref{marginal scores} presents the frequencies of numbers of goals scored by the home and away team. It shows, in particular, that the home team scored at least 1 goal in 76.6\% of the matches and the away team scored more than 1 goal in only 24.6\%.

In this data set, the minute in which a goal is scored or a red card is received are discrete variables. Therefore, we can compute both the goal and the red card rates in each minute dividing the total number of goals (red cards) scored in a specific minute by the total number of matches.

\begin{figure} [ht] 
\caption{Goal rates (excluding the stoppage times)}
\label{goal rate} 
\centering
\includegraphics[width=0.95\textwidth]{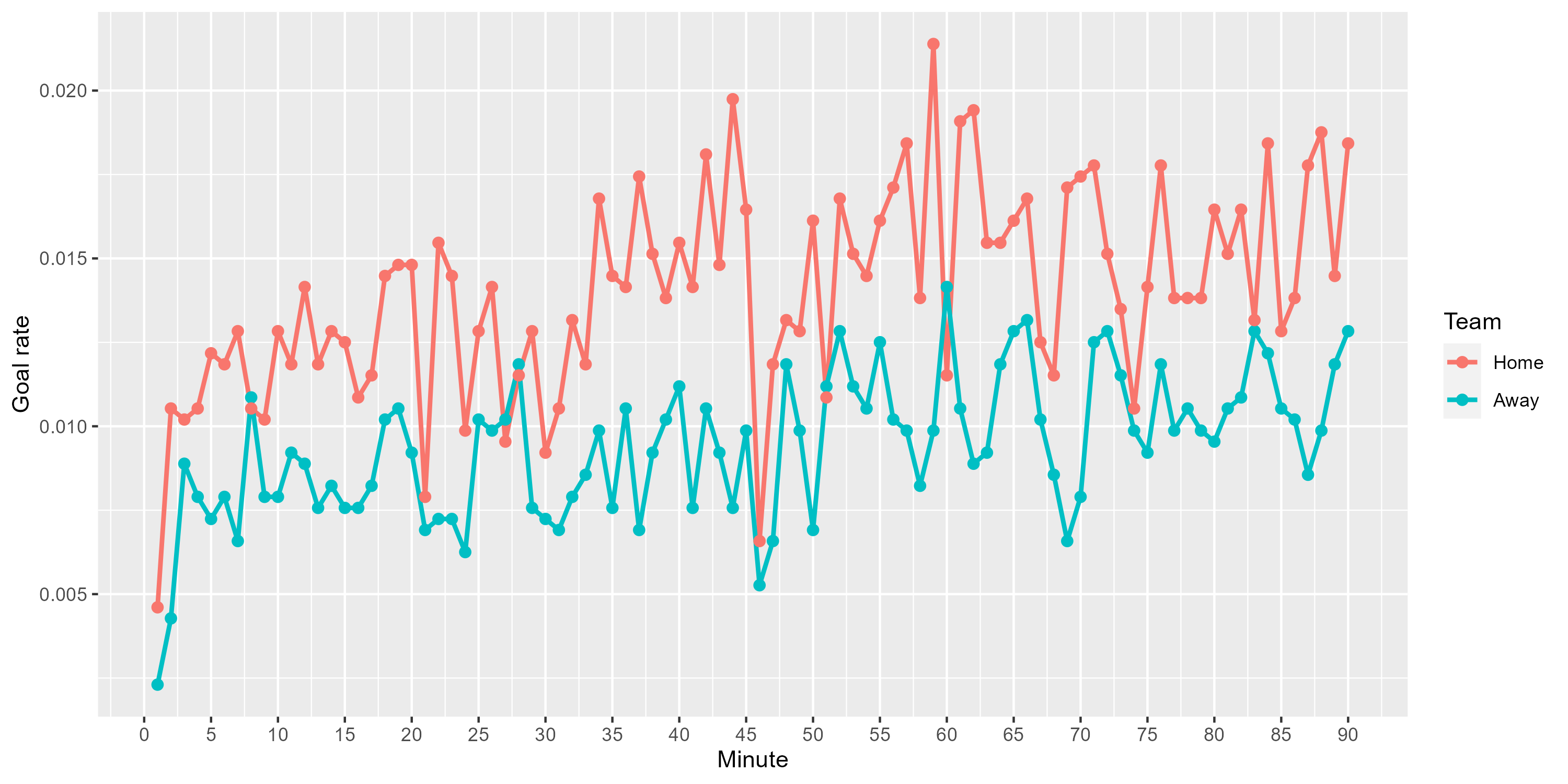}
\end{figure}

The goal rate for the regular 90 minutes (excluding extra time) is shown in Figure \ref{goal rate}. It can be clearly seen this rate is greater in the second half than in the first half and it tends to increase over time. This fact may be explained by the fatigue of the players and the reduction of the time available to score a goal, which lead teams that are losing to take more risks. 

\begin{figure} [ht] 
\caption{Red card rates excluding the stoppage times}
\label{fig:red_card_rate} 
\centering
\includegraphics[width=0.95\textwidth]{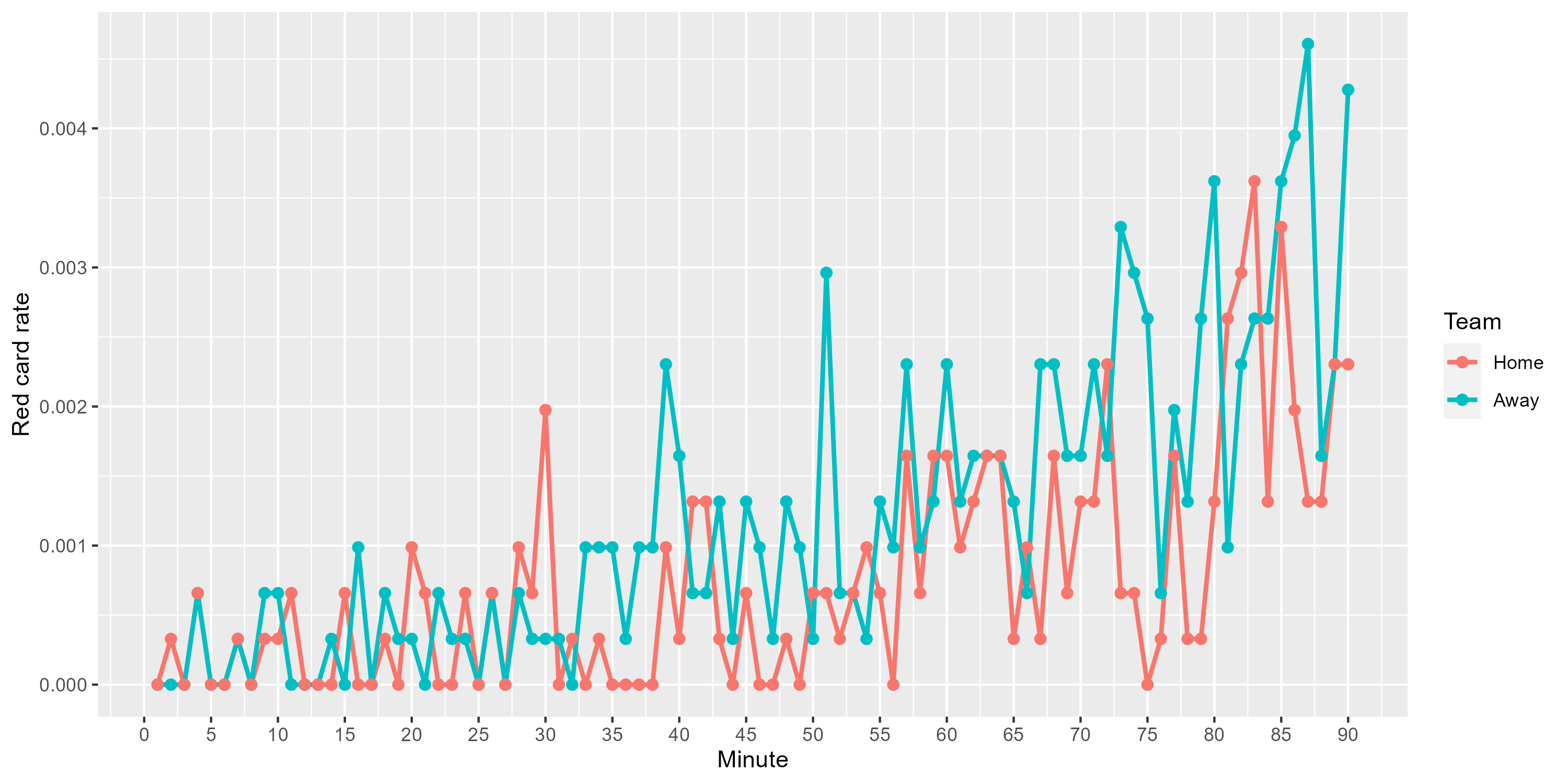}
\end{figure}

Similarly to the goals, the red card rate is shown in Figure \ref{fig:red_card_rate}. This rate is also higher in the second half and it also tends to increase over time. A possible explanation for this phenomenon is that a possible way for a player to be sent off is to receive two yellow cards, which usually takes some time to happen.

\begin{figure} [ht] 
\caption{Frequency of the goal difference (Home $-$ Away) for matches with/without red cards for the away team}
\label{fig:with_away_red_card} 
\centering
\includegraphics[width=0.95\textwidth]{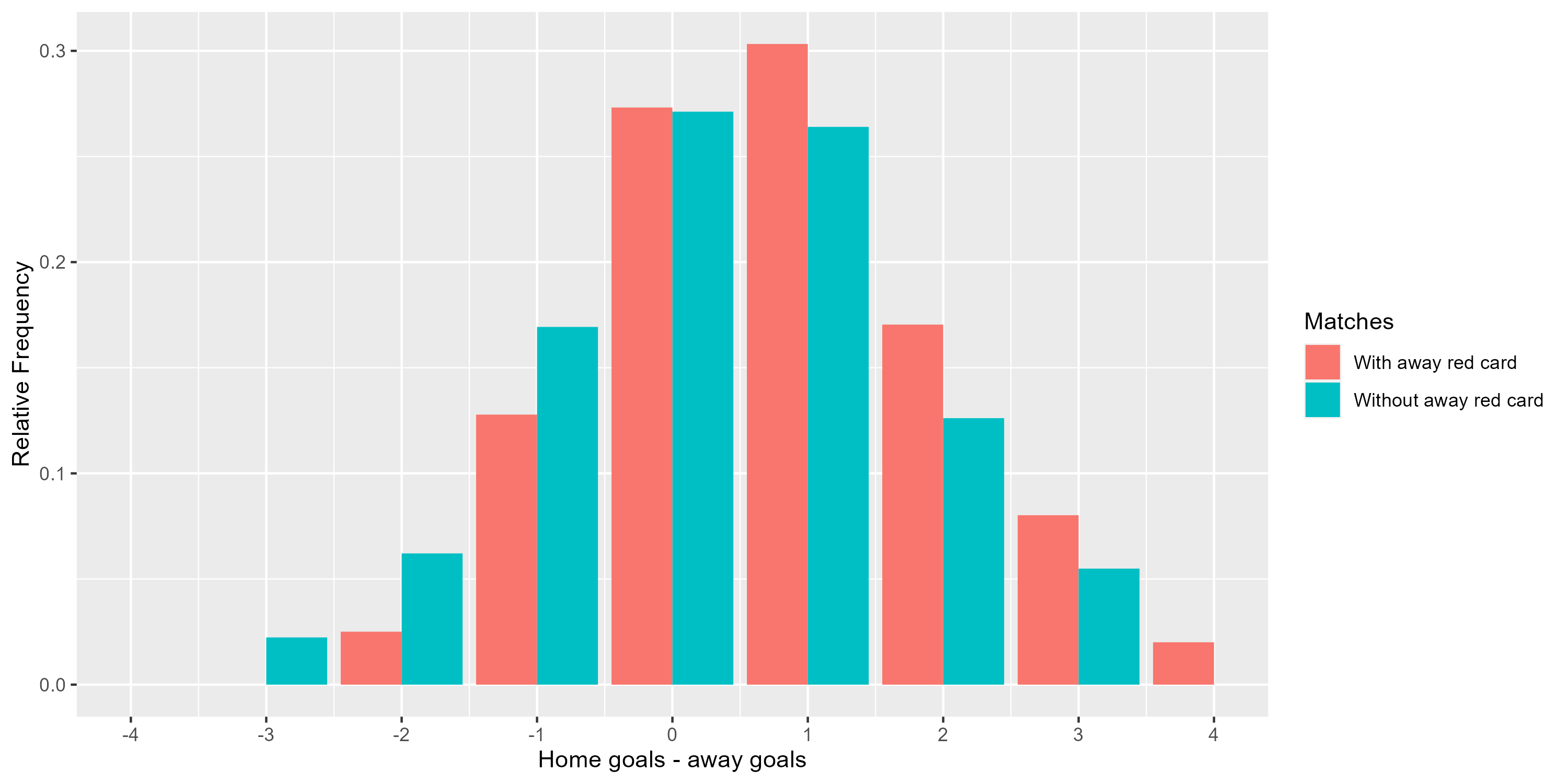}
\end{figure}

\begin{figure} [ht] 
\caption{Frequency of the goal difference (Home $-$ Away) for matches with/without red cards for the home team}
\label{fig:with_home_red_card} 
\centering
\includegraphics[width=0.95\textwidth]{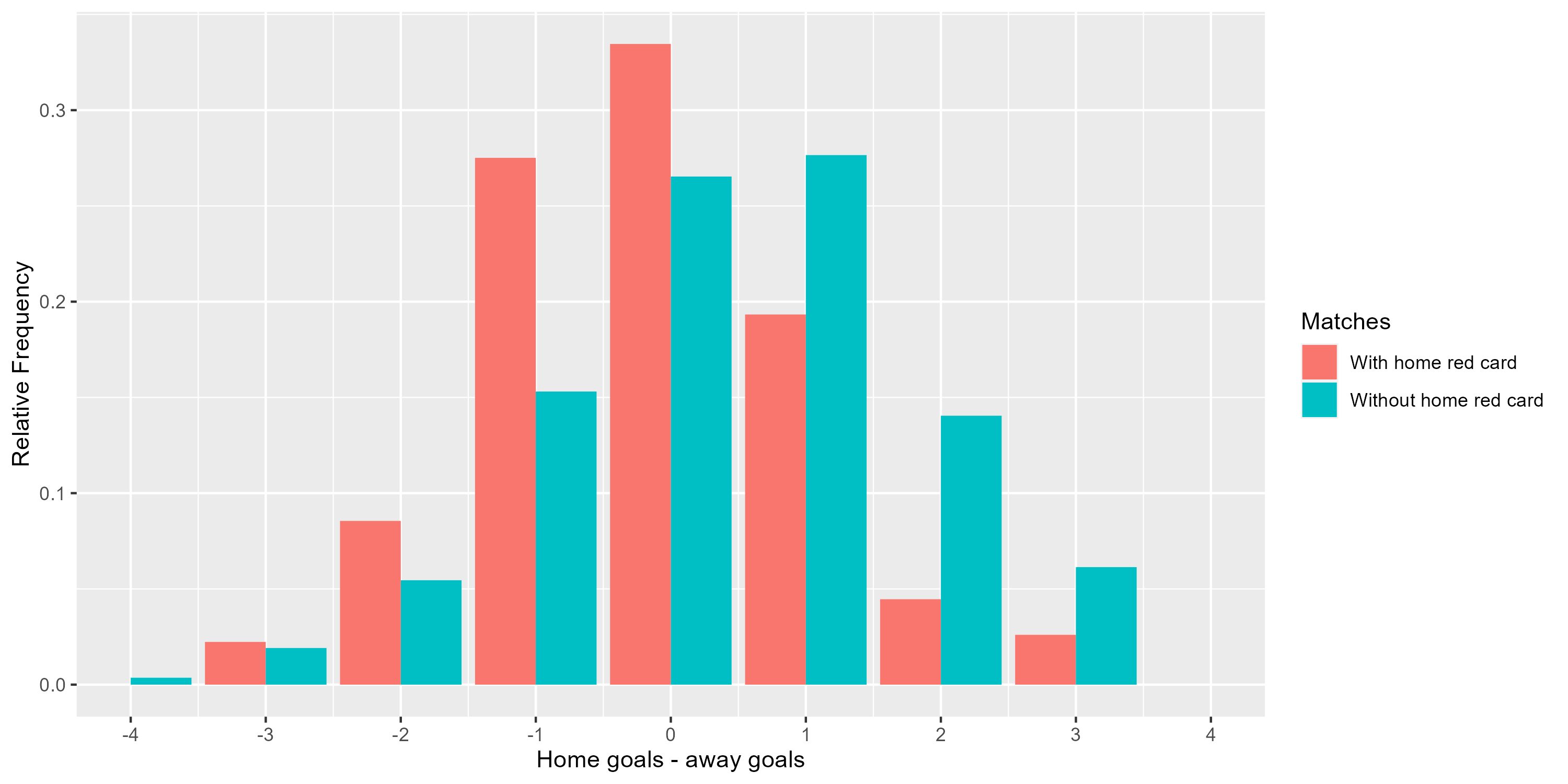}
\end{figure}

Figures \ref{fig:with_away_red_card} and \ref{fig:with_home_red_card} plot the histograms of the goal difference (Home goals $-$ Away goals) for matches with and without red cards. Figure~\ref{fig:with_away_red_card} considers red cards for the away team, while \ref{fig:with_home_red_card} presents the  histograms considering red cards for the home team. As one might expect, a red card for a team gives an advantage to the opposing team which is reflected in the goal differences.

At the end of each regular 45 minutes of a half, the referee may decide to extend its length by adding {\em stoppage time} (aka injury time) to compensate for the interruptions that occurred during the regular 45 minutes due to injuries or other events.

\begin{figure} [ht] 
\caption{Stoppage time distribution for the first and second halves. }
\label{fig:stoppage_time} 
\centering
\includegraphics[width=0.95\textwidth]{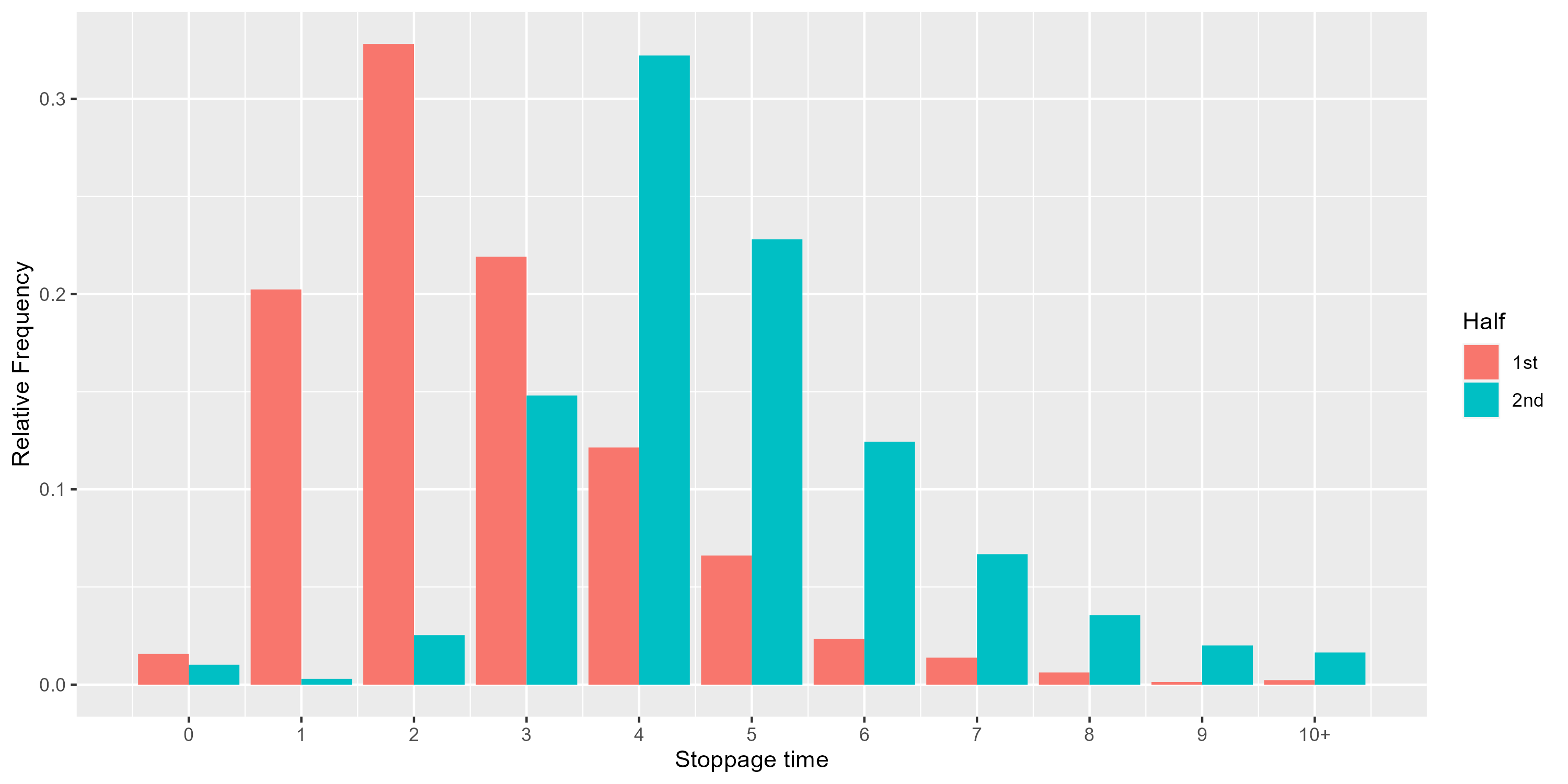}
\end{figure}

Figure \ref{fig:stoppage_time} displays the histograms of stoppage time in the first and second halves. The stoppage time in the second half is likely to be longer than in the first half. This may be due to the greater number of interruptions in the second half, mainly caused by players substitutions. 

As some stoppage time is usually added to the regular 45 minutes halves (see Figure \ref{fig:stoppage_time}), there is always a chance that goals will be scored during the additional time. In fact, 3.4\% (6.7\%) of the goals have been scored  during the additional minutes of the first (second) half.

\begin{figure} [ht] 
\caption{Difference between (a) the logarithm of the value of the teams and (b) goals scored by the home/away teams}
\label{fig:market_value_diff_goals} 
\centering
\includegraphics[width=0.95\textwidth]{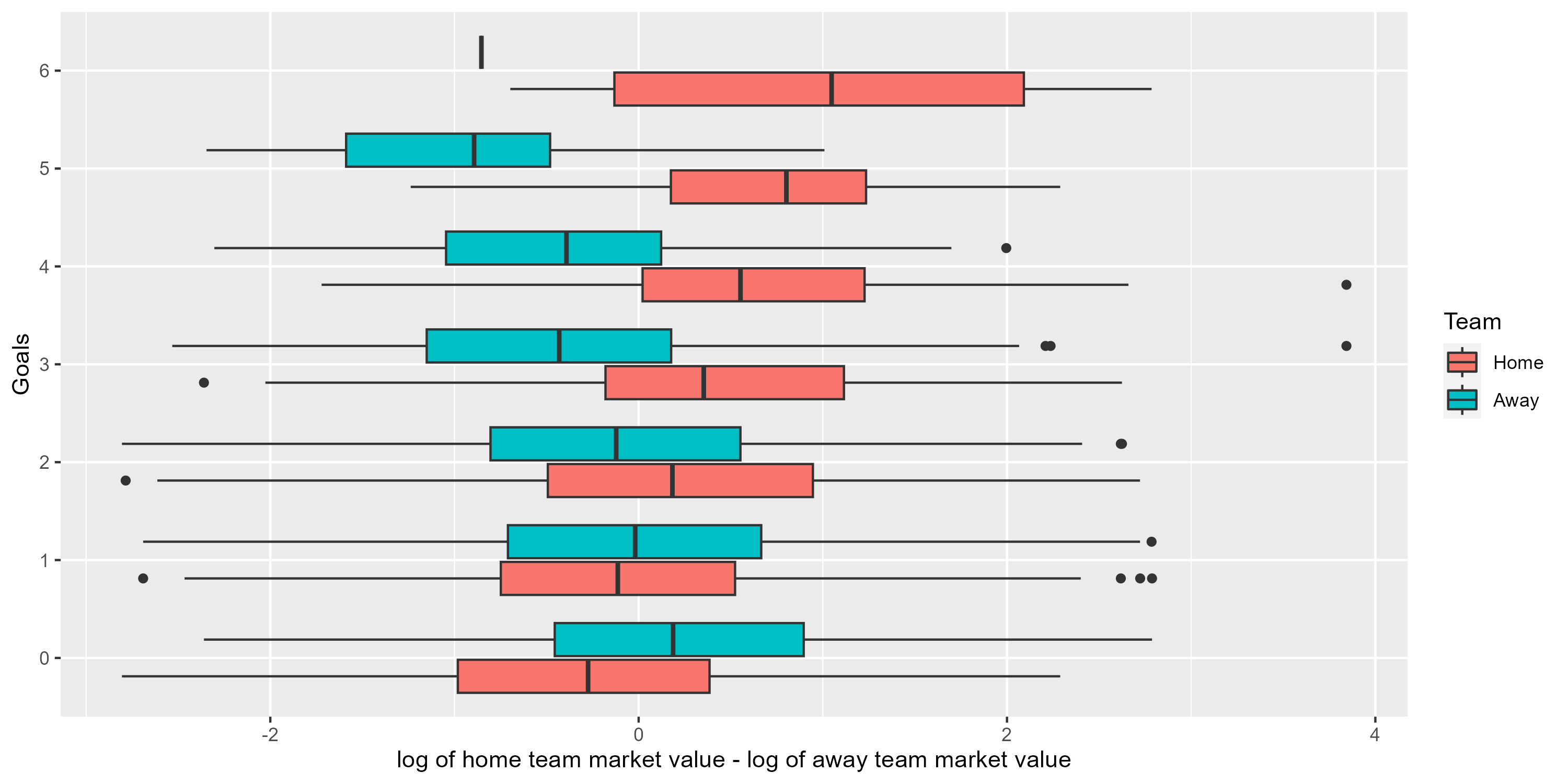}
\end{figure}

The mean market value of the starting players of each team is present in Table \ref{tab:value}. Flamengo and Palmeiras stand out as having the most valuable players in the league. They won 5 out of the 8 seasons in this data set. Figure~\ref{fig:market_value_diff_goals} illustrates the statistical dependence of the numbers of goals scored an the relative difference in the market values. 
It can be seen that, as the home team's market value relative to the away team's value increases, the home team tends to score more while the away teams scores less.
Accordingly, teams with higher market value tend to a better placement in the league; see Table \ref{tab:value}.

\begin{table}[ht]
\centering
\caption{Mean market value of the starting 11 players by team in million of euros, number of seasons played and mean position in the league.}
\label{tab:value}
\begin{tabular}{rlrrr}
\hline
& Team & Mean market value & Seasons & Mean position \\ 
\hline
1 & Flamengo & 49.08 &   8 & 4.00 \\ 
2 & Palmeiras & 43.27 &   8 & 3.38 \\ 
3 & Grêmio & 34.31 &   7 & 6.71 \\ 
4 & Atlético-MG & 33.15 &   8 & 5.62 \\ 
5 & São Paulo & 33.05 &   8 & 8.00 \\ 
6 & Corinthians & 32.17 &   8 & 6.38 \\ 
7 & Red Bull Bragantino & 29.07 &   3 & 10.00 \\ 
8 & Santos & 28.06 &   8 & 6.75 \\ 
9 & Internacional & 26.61 &   7 & 6.86 \\ 
10 & Cruzeiro & 26.42 &   5 & 10.00 \\ 
11 & Fluminense & 22.14 &   8 & 10.12 \\ 
12 & Athletico-PR & 18.91 &   8 & 8.50 \\ 
13 & Vasco & 14.11 &   5 & 14.00 \\ 
14 & Botafogo & 13.45 &   6 & 11.67 \\ 
15 & Sport & 13.31 &   6 & 14.50 \\ 
16 & Ponte Preta & 11.60 &   3 & 12.67 \\ 
17 & Figueirense & 11.57 &   2 & 17.00 \\ 
18 & Bahia & 11.53 &   5 & 13.20 \\ 
19 & Vitória & 11.13 &   3 & 17.00 \\ 
20 & Chapecoense & 9.98 &   6 & 14.33 \\ 
21 & Fortaleza & 9.88 &   4 & 9.25 \\ 
22 & Coritiba & 9.57 &   5 & 16.20 \\ 
23 & Goiás & 9.19 &   4 & 15.00 \\ 
24 & Ceará & 8.16 &   5 & 14.00 \\ 
25 & Cuiabá & 8.11 &   2 & 15.50 \\ 
26 & CSA & 8.07 &   1 & 18.00 \\ 
27 & Atlético-GO & 7.72 &   4 & 15.00 \\ 
28 & Joinville & 7.49 &   1 & 20.00 \\ 
29 & América-MG & 7.20 &   4 & 13.75 \\ 
30 & Santa Cruz & 7.09 &   1 & 19.00 \\ 
31 & Avaí & 7.05 &   4 & 18.50 \\ 
32 & Juventude & 6.53 &   2 & 18.00 \\ 
33 & Paraná & 5.08 &   1 & 20.00 \\  
\hline
\end{tabular}
\end{table}

\section{The model} \label{sec:notation}

This section describes a generic model for the joint dynamics of ``relevant events" in a football match. In addition to goals and red cards discussed in Section~\ref{sec:data}, the events could also include penalty shots, injuries, etc. We start by presenting the event model conditionally on the total length of a match. The total length depends on the stoppage times whose distribution may depend on the events in the preceding half. The full model is then obtained by combining the event model with the stoppage time model.

\subsection{Event processes}

The set of relevant events will be denoted by $J$ and it is assumed to be finite. We assume that the occurrence times of each $j\in J$ follows a {\em Cox process} (also known as doubly stochastic Poisson processes); see e.g.\ \cite{kal2}. That is, each $j\in J$ has an intensity process $\lambda^j$ such that, conditionally on $\lambda^j$, the number of events follows an inhomogeneous Poisson processes with intensity $\lambda^j$. This means that, conditionally on $\lambda^j$, the first event time $t$ after a given time $s\ge 0$ has cumulative distribution function
\[
\Phi^j_{\lambda^j}(s;t) = 1-\exp(-\Lambda^j(s,t)),
\]
and density
\[
\phi^j_{\lambda^j}(s;t) = \frac{d\Phi^j_{\lambda^j}(s;t)}{dt} =\lambda^j(t)\exp(-\Lambda^j(s,t)),
\]
where
\[
\Lambda^j(s,t):=\int_s^t\lambda^j(r)dr.
\]

Denote the total length of a match by $T$ and let $t^j_l\in[0,T]$ be the time of occurrence of the $l$th event of type $j\in J$. Let $L^j$ be the total number of events of type $j\in J$ during the game. We assume that, conditionally on the length $T$ of the game and on the intensities $\lambda = (\lambda^1, \ldots, \lambda^J)$, the processes $J$ are independent of each other. It follows that the likelihood of observing event times $t^j_l$, $l=1,\ldots,L^j$, $j\in J$ is given by
\begin{align*}
{\cal L}(\lambda\mid (t_1^j,\ldots, t_{L^j}^j)_{j\in J},T) &= \prod_{j\in J}\prod_{l=1}^{L^j}\phi^j_{\lambda^j}(t^j_{l-1};t^j_l)\exp(-\Lambda^j(t^j_{L^j},T))\\
&=\prod_{j\in J} \prod_{l=1}^{L^j} \lambda^j(t^j_{t_l})\exp(-\Lambda^j(t^j_{l-1},t^j_l))\exp(-\Lambda^j(t^j_{L^j},T))\\
&= \prod_{j\in J}\exp(-\Lambda^j(0,T)) \prod_{l=1}^{L^j}\lambda^j(t^j_l)
\end{align*}
where $t^j_0:=0$. The exponential terms $\exp(-\Lambda^j(t^j_{L^j},T))$ above are the probabilities of the match ending before the $(L^j+1)-$th event of type $j$ occurs.

\subsection{Stoppage time}\label{sec:et}

The total duration $T$ of a match is the sum of the durations of the two halves of a match. The total length of a game is thus
\[
T=45+U_1+45+U_2,
\]
where $U_k$ is the length of the stoppage time of the $k$th half.

The stoppage times come in multiples of minutes and they are announced after the regular 45 minutes of each half. We will assume that $U_k$ follows a Poisson distribution with a parameter $\pi_k$ that may depend on the events observed by the 45th minute of the $k$th half. The likelihood function for the length $U_k$ is
\[
\L(\pi_k\mid U_k) = \frac{\pi^{U_k} e^{-\pi_k}}{U_k!}.
\]

\subsection{The full likelihood}\label{sec:fl}

Assuming that, conditionally on the intensities $\lambda$ and the Poisson parameters $\pi_k$, the events across the two halves of a game are independent, the likelihood function for the events throughout the game becomes
\begin{align}
\L(\lambda,\pi_1,\pi_2\mid (t_1^j,\ldots, t_{L^j}^j)_{j\in J},T) &= \L(\lambda, T\mid (t_1^j,\ldots, t_{L^j}^j)_{j\in J})\prod_{k=1,2}\L(\pi_k\mid U_k)\nonumber\\
&= \prod_{k=1,2}\frac{\pi^{U_k}e^{-\pi_k}}{U_k!}\times \prod_{j\in J}\exp(-\Lambda^j(0,T)) \prod_{l=1}^{L^j}\lambda^j(t^j_l),\label{fullL}
\end{align}
where, again, $T=90+U_1+U_2$.

If we have data over several matches, and if we assume that, conditionally on the intensities, the events are independent across the matches, then the overall likelihood function is simply the product of likelihood functions of single matches. The intensities may, of course, vary from match to match.

\section{Parameterization of the intensities}\label{sec:params}

In order to make the specification of the model implementable, we will parameterize the intensity processes and the Poisson parameters of the stoppage times by a finite set of real parameters. This section describes a general format of parameterizations that is both general as well as computationally tractable in the sense that the resulting log-likelihood functions are strictly concave. The concavity allows us to apply the general theory and software of convex optimization to compute maximum likelihood estimates of the parameters.

\subsection{Event processes}

We will parameterize the intensity processes $\lambda^j$ so that their logarithms are linear functions of the parameters. That is, we set
\begin{equation}\label{xipsi}
\ln\lambda^j(t) = \sum_{i\in I^j}\xi^j_i\psi^j_i(t),
\end{equation}
where $\{\psi^j_i\mid i\in I^j$\} is a finite set of regressors for the intensity of event type $j\in J$ and the $\xi^j_i$ are the corresponding parameters. The above parameterization results in a (strictly) concave log-likelihood function (see Theorem~\ref{thm:concave} below), which facilitates the estimation of the parameters. The format is also very general as it covers most existing football models and many more. 

In the applications of the following sections, we will assume that the regressors $\psi^j_i$ are stochastic processes that are adapted to the available information. That is, the values of $\psi^j_i(t)$ only depend on events observed by time $t$. This is important when the model is used to describe the conditional distribution of future events at a given point in time.

For example, classical static score models (which do not consider stoppage time) studied e.g.\ in \cite{maher1982modelling,dixon1997modelling}, correspond to constant intensity processes that only depend on the teams in the game. In such models, the total number of goals for the two teams are independent and Poisson distributed; see Section~\ref{ssec:sgoalmodels} below. The dynamic models of Dixon and Robinson~\cite{dixon1998birth}, include processes $\psi^j_i$ which are functions of the current score or simply deterministic functions of time. The generic parameterization \eqref{xipsi} above allows the intensities to depend also on e.g.\ the number of red cards given by time $t$; see Section~\ref{sec:sr} below.

\subsection{Stoppage time}

To specify the full likelihood, we need to parameterize the distribution of the length of the stoppage time; see Section~\ref{sec:fl}. As in Section~\ref{sec:et}, we assume the length of the stoppage time at the end of the $k$th half is Poisson distributed with parameter $\pi_k$. We will assume that
\begin{equation}\label{eq:pi}
\ln\pi_k = \sum_{i\in I^k_s}\xi_i\phi_i,
\end{equation}
where $\{\phi_i\mid i\in I^k_s\}$ is a collection of regressors whose values are observable at the beginning of the stoppage time.

\subsection{Full likelihood}\label{sec:full-likelihood}

The full log-likelihood function in terms of all the parameters 
\[
\xi = \left( ((\xi_i^j)_{i\in I^j} )_{j\in J} , ((\xi^k_i)_{i \in I^k_s})_{k=1,2} \right)
\] 
is obtained by substituting the expressions for $\lambda^j$ and $\pi^k$ from \eqref{xipsi} and \eqref{eq:pi} to the full likelihood function \eqref{fullL}. This gives the parametric likelihood function
\begin{align*}
\ell(\xi \mid (t_1^j,\ldots, t_{L^j}^j)_{j\in J},T) =& -\sum_{j\in J} \left[\int_0^{T}\exp\left(\sum_{i\in I^j}\xi^j_i\psi^j_i(r)\right)dr - \sum_{l=1}^{L^j}
\left(\sum_{i\in I^j}\xi^j_i\psi^j_i(t^j_l)\right)\right]\\
&+\sum_{k=1,2}\left[U^k\sum_{i\in I^k_s}\xi_i\phi_i-\exp\left(\sum_{i\in I^k_s}\xi_i\phi_i\right)-\ln(U^k!)\right].
\end{align*}
When we have data over several matches, the full log-likelihood function is simply the sum of the likelihood functions of all the matches.

The following establishes the concavity of the log-likelihood function as a function of the parameters. Concavity makes the numerical solution of the maximum likelihood estimation problem computationally  robust and fast. When the regressors are linearly independent, we find that the log-likelihood function is strictly convex. This implies that the maximum likelihood estimates are unique. 

We denote the set of all parameters in the model by
\[
I:=\bigcup_{j\in J}I^j\cup \bigcup_{k=1,2}I^k_s.
\]

\begin{theorem}\label{thm:concave}
The log-likelihood function $\ell$ is concave on $\reals^I$. If the regressors are linearly independent on the given data, then $\ell$ is strictly concave.
\end{theorem}

\begin{proof}
For a single match, the negative $\ell$ of the log-likelihood function is the sum of the functions 
\[
\ell^j(\xi^j):=\int_0^T\exp\left(\sum_{i\in I^j}\xi^j_i\psi^j_i(r)\right)dr - \sum_{l=1}^{L^j}
\left(\sum_{i\in I^j}\xi^j_i\psi^j_i(t^j_l)\right)
\]
over event types $j\in J$ and of the functions
\[
\ell^k_s(\xi_s):=-U^k\ln\left(\sum_{i\in I^k_s}\xi_i\phi_i\right) + \sum_{i\in I^k_s}\xi_i\phi_i + \ln(U^k!)
\]
over the two halfs $k=1,2$. The full likelihood function is the sum over the matchwise likelihood functions. To prove the convexity of the sum, it suffices to show that each term is convex. Strict convexity then follows if any of the functions in the sum is strictly convex.

As for the event types, the log-intensities $\eta^j:=\ln \lambda^j$ are elements of the linear space $\S$ of $\reals$-valued $\F\otimes\B(\reals_+)$-measurable functions on $\Omega\times\reals_+$. Here $\B(\reals_+)$ denotes the Borel sigma-algebra on $\reals_+$. The function $\ell^j$ is the composition of the linear mapping $\Psi^j:\reals^{I^j}\mapsto\S$ given by
\[
\Psi^j(\xi):= \sum_{i\in I^j}\xi^j_i\psi^j_i
\]
and of the function 
\begin{align*}
l^j(\eta^j) = \int_0^T\exp\left(\eta^j(r)\right)dr - \sum_{l=1}^{L^j}
\eta^j(t^j_l)
\end{align*}
which is strictly convex on $\S$. As to the the stoppage times, the function $\ell_s^k$ is the composition of the linear mapping
\[
\xi_s\mapsto \sum_{i\in I^k_s}\xi_i\phi_i
\]
and the strictly convex function
\[
\eta\mapsto -U^k\ln\eta + \eta + \ln(U^k!)
\]
which is strictly convex function on $\reals$.

The first claim now follows from the fact that compositions of linear mappings and convex functions are convex. The second claim follows from the fact that the composition of an injective linear mapping with a strictly convex function is strictly convex. The injectivity means that the regressors are linearly independent on the data.
\end{proof}

Other than being adapted to the available information, we haven't said anything about the regressors $\psi^j_i$ and $\phi_i$ yet. In the examples studied in the following sections, the regressors may depend not only on time and the events within a given game but they also depend on the season and, not surprisingly, the teams playing in a match. The corresponding basis functions may fail to be linearly independent, as happens e.g.\ in the classical model of \cite{maher1982modelling}; see Section~\ref{ssec:sgoalmodels} below. Under linear dependence, the maximizer of the likelihood function is not unique. The uniqueness can be achieved by imposing linear equality constraints on the parameters. The maximum likelihood estimation then becomes a convex optimization problem with linear constraints. Linear equality constraints are still within the realm of convex optimization and easily handled by standard convex optimization packages.

\section{Specifying the regressors}\label{sec:sr}

This section gives examples of regressors for different event types. We will consider the events
\[
J := \{\text{home goals, away goals, home red cards, away red cards}\}.
\]
We start with the classic model of \cite{maher1982modelling} that describes the final score distributions by two independent Poisson variables. Such a model corresponds to constant intensities in the framework of Section~\ref{sec:notation}. Still in the static setting, we then introduce new regressors that depend on the values of the teams' players at the beginning of the game. The remaining part of the section is concerned with dynamic regressors $\psi^j_i$ that may depend on events that occur during the game.

\subsection{Static regressors for goals}\label{ssec:sgoalmodels}

The classical static score models studied e.g.\ in \cite{maher1982modelling} and \cite{dixon1997modelling} correspond to constant intensity processes that only depend on the teams in the game. In such models, the total number of goals are Poisson distributed with parameters $\alpha_h\beta_a\gamma$ for the home team and $\alpha_a\beta_h$ for the away team. Here $\alpha_h$ is the ``attack parameter" of the home team, $\beta_a$ is the ``defense parameter" of the away team and $\gamma$ is the ``home advantage parameter", whose significance has been also studied in \cite{home_clarke_1995, home_pollard_2008, 
homefield_carmichael_2005}.

Such models can be specified by choosing $I=K\times\{h,a\}$, where $K$ is the set of teams in the league, and setting
\[
\psi^j_{k,h}=\begin{cases}
1 & \text{if $k$ is the home team},\\
0 & \text{otherwise}.
\end{cases}
\]
and
\[
\psi^j_{k,a}=\begin{cases}
1 & \text{if $k$ is the away team},\\
0 & \text{otherwise}.
\end{cases}
\]
If $j\in J$ is home goals, then $\xi^j_{k_h,h}=\ln\alpha_{k_h}$ and $\xi^j_{k_a,a}=\ln\beta_{k_a}$. Here $k_h,k_a\in K$ denote the home and away teams in a given game. The home advantage factor $\gamma$ is simply an additive constant term in the home goal intensity process, or equivalently, the coefficient of the constant regressor $\psi^j_i=1$.

As such the model is over parameterized as we get the same intensities if we multiply each attack parameter by a constant and divide the defence parameters by the same constant. Dixon and Coles chose to restrict the parameters so that the average of the attack parameters over all the teams equals one. In the computational experiments below, we will use the restriction that the geometric averages of the attack parameters equals that of the defence parameters. Such a constraint can be written as
\[
\sum_{k\in K}\ln\alpha_k = \sum_{k\in K}\ln\beta_k.
\]
This has the benefit of making the constrained log-likelihood maximization problem a convex optimization problem and thus, amenable to efficient convex optimization solvers; see Section~\ref{sec:fitting} below. Indeed, $\ln\alpha_k$ and $\ln\beta_k$ correspond to some of the parameters $\xi^ j_i$ in the general formulation in Section~\ref{sec:full-likelihood} so the above constraint is linear. Linear equality constraints are readily handled by standard convex optimization solvers. The above constraint also makes the attack and defence parameters of comparable scale; see Section~\ref{ssec:goalstats} below.

It is reasonable to assume that the market values of the players of the teams is reflected in the goal intensities. We will denote by $V^k$ the market value of the 11 players of the starting line-up of team $k$. We model the dependence of the intensities on the market values by including the regressors
\[
\psi^j_i = \ln V^h - \ln V^a
\]
if $j=``\text{home goals}"$ and 
\[
\psi^j_i = \ln V^a - \ln V^h
\]
if $j=``\text{away goals}"$. Obviously, we could use the same regressor for both event types by simply changing the sign of the corresponding multiplier. The above regressors correspond to the hypothesis that the log-intensities depend linearly on the logarithm of the ratio of the values of the two teams.

\subsection{Dynamic regressors for goals}\label{ssec:dgoalmodels}

\cite{dixon1998birth} as well as \cite{tcrg} proposed dynamic intensity models where the goal intensities during the game depend on the current score through ``dummy variables" that describe certain sets of scores. We will denote by $N^h(t)$ and $N^a(t)$ the home and away scores, respectively, at time~$t$. We will assume that the log-intensities for goals depend linearly on the goal difference by using the regressors
\[
\psi^j_i(t) = N^h(t)-N^a(t),
\]
if $j=``\text{home goals}"$ and 
\[
\psi^j_i(t) = N^a(t)-N^h(t)
\]
if $j=``\text{away goals}"$.

Possible red cards during a game reduce the number of players in the receiving team so it is natural to assume that the red cards affect the goal intensities. We denote by $N^{hr}(t)$ and $N^{ar}(t)$ the numbers red cards received by the home team and the away team, respectively, by time $t$. We describe the dependence by the regressor
\[
\psi^j_i(t) = N^{ar}(t)-N^{hr}(t)
\]
if $j=``\text{home goals}"$ and 
\[
\psi^j_i(t) = N^{hr}(t)-N^{ar}(t)
\]
if $j=``\text{away goals}"$.
Note that when this covariate is added to a model it generates a dependence between home and away scores, since when one team scores it alters its own rate as well as the opponent's.

The last regressor is a simple half time indicator
\[
\psi^j_i(t) = 1_{\{t\ge 45+u\}}
\]
which models the fact that goal intensities seem to be different in the two halves of a match.

\subsection{Regressors for stoppage time} \label{ssec:regressors_stoppage_time}

It is reasonable to assume that the length of the stoppage time depends on the interruptions and other events during the played half. For example, it may depend on the number of goals scored as the players tend to take some time for celebrations, and also because, since 2019, all goals must be checked by Video Assistant Referee (VAR). The length of the stoppage time may also depend on the number of red cards as they are also checked by VAR. Red cards can also be associated with injuries that usually require time for medical care.

Accordingly, we will regress the log-parameter of the Poisson-distributed stoppage time in Section~\ref{sec:et} on the regressors
\[
\phi_i=\begin{cases} N^h(45)+N^a(45), \text{ in the first half;} \\ N^h(90+U_1)+N^a(90+U_1) - N^h(45+U_1)-N^a(45+U_2), \text{ in the second half}
\end{cases}
\]
and 
\[
\phi_i=\begin{cases} N^h(45)+N^a(45), \text{ in the first half;} \\ N^{hr}(90+U_1)+N^{ar}(90+U_1) - N^{hr}(45+U_1)-N^{ar}(45+U_2), \text{ in the second half}.
\end{cases}
\]
where $t_e$ denotes the end of the regular time of a given half. We will also include a constant term, i.e.\ a regressor of the form
\[
\phi_i = 1.
\]

Another factor that influences stoppage time, mainly during the second half, is the current score of the match. When one of the teams is in a significant lead, the stoppage time tends to be short as the winner is already largely decided. We model this effect by adding the regressor
\[
\phi_i = 1_{\{|N^h(90+U_1)-N^a(90+U_1)|\le 1\}}
\]
to the log-Poisson parameter of the stoppage time of the second half. A nonzero multiplier for this regressor would mean that the length of the stoppage time depends on whether the score difference at the end of the half is more than $1$ or not.

\subsection{Regressors for red cards}\label{sec:red-cards}

As discussed in Section \ref{sec:data}, Figure \ref{fig:red_card_rate}, historically, rates for red cards tend to increase with the development of the matches. To model this, we use the basis functions
\[
\psi^j_i(t) = 1
\]
and
\[
\psi^j_i(t) = \ln t.
\]
Note that this implies that the red card intensity has the form
\[
\lambda^j(t) = e^{\xi^j_{i_0}}t^{\xi^j_{i_1}},
\]
where $i_0$ and $i_1$ are the indices for the constant regressor and the logarithmic regressor, respectively.

\section{Fitting the model for Brazilian Serie A}\label{sec:fitting}

This section fits the model presented in Sections~\ref{sec:notation} and \ref{sec:sr} to the data on  Campeonato Brasileiro de Futebol Série A described in Section~\ref{sec:data}. We will study the explanatory power of the regressors defined in Section~\ref{sec:params}. 

By Theorem~\ref{thm:concave}, the resulting log-likelihood function is convex which makes the maximum likelihood estimation of the parameters $\xi$ a problem of convex optimization. This allows us to  find globally optimal estimates using standard techniques of convex optimization. If the regressors are linearly independent, the maximum likelihood estimate will be unique, by Theorem~\ref{thm:concave}. 

The numerical optimization was done using the primal-dual interior point solver of MOSEK \cite{mosek}. The maximum likelihood estimation problem was formulated and communicated to the solver using CVXR package \cite{CVXR} in the R programming language \cite{rstats}.

\subsection{Goal models}\label{ssec:goalstats}

We study the explanatory power of the different regressors for the goal intensities by fitting five different models to the data (denoted G0$-$G4), summarized in Tables \ref{tab:home_regressors} and \ref{tab:away_regressors}. A complete model for goals should be understood as the combination of a model for home and a model for away goals.

The models are nested with covariates described in Section \ref{sec:sr}. Model G0 is simply the static Maher model with attack, defence and home advantage factors only. Model G1 adds the teams' value regressor, Model G2 adds the half time regressor, Model G3 adds the goals and model G4 adds the red cards. 

For simplicity, we will impose the restriction that the parameters of the regressors are the same for the home and away goal processes, with the exception of the home advantage parameter which is set to zero for the away team. Allowing for arbitrary parameters did not improve the model fit. Table \ref{tab:away_regressors} presents a summary similar to the one on Table \ref{tab:home_regressors}.

\tiny
\begin{table}[h]
    \centering
    \begin{tabular}{cccccccc}
    \toprule
         Model & \text{Attack} & \text{Defence} & \text{Home} & \text{Values} & \text{Half time} & \text{Goals} & \text{Red cards} \\
    \midrule
    G0 & \ding{51} & \ding{51} & \ding{51} &   &   &   &   \\
    G1 & \ding{51} & \ding{51} & \ding{51} & \ding{51} &   &   &   \\
    G2 & \ding{51} & \ding{51} & \ding{51} & \ding{51} & \ding{51} &   &   \\
    G3 & \ding{51} & \ding{51} & \ding{51} & \ding{51} & \ding{51} & \ding{51} &   \\
    G4 & \ding{51} & \ding{51} & \ding{51} & \ding{51} & \ding{51} & \ding{51} & \ding{51} \\
    \bottomrule
    \end{tabular}
    \caption{Home goal regressors}
    \label{tab:home_regressors}
\end{table}
\normalsize

\tiny
\begin{table}[h]
    \centering
    \begin{tabular}{cccccccc}
    \toprule
                  Model & \text{Attack} & \text{Defence} & \text{Home} & \text{Values} & \text{Half time} & \text{Goals} & \text{Red cards} \\
    \midrule
    G0 & \ding{51} & \ding{51} &   &   &   &   &   \\
    G1 & \ding{51} & \ding{51} &   & \ding{51} &   &   &   \\
    G2 & \ding{51} & \ding{51} &   & \ding{51} & \ding{51} &   &   \\
    G3 & \ding{51} & \ding{51} &   & \ding{51} & \ding{51} & \ding{51} &   \\
    G4 & \ding{51} & \ding{51} &   & \ding{51} & \ding{51} & \ding{51} & \ding{51} \\
    \bottomrule
    \end{tabular}
    \caption{Away goal regressors}
    \label{tab:away_regressors}
\end{table}
\normalsize

Table \ref{tab:aic_goals} presents a comparison between the five different models (which include home and away goals). As the model are nested, the log-likelihoods are strictly increasing when moving from model G0 (the least flexible one) to G4. The increase in the log-likelihood comes with the price of an increased number of parameters, from 67 parameters in model G0 to 71 parameters in model G1. In model G0, the parameters are the attack and defence parameters for each one of the 33 teams in our data set (see Table~\ref{tab:value}) plus the home advantage parameter. As described in Tables \ref{tab:home_regressors} and \ref{tab:away_regressors}, each model from G1 to G4 adds one extra parameter, leading to model G4 having 71 parameters. 

The fourth and fifth columns of Table \ref{tab:aic_goals} present the Akaike Information Criterion (AIC) and Bayesian Information Criterion (BIC), respectively. Even though the extra parameters are penalized by both AIC and BIC, they are not enough to outweigh the increased likelihood when moving from model G0 to G4 which is preferred by both criteria. The last column presents the p-value of the likelihood ratio test, always comparing with the previous model. Since all values are pretty close to zero, it points, again, that more complex models are better.

\begin{table}[ht]
\centering
\begin{tabular}{cccccc}
\toprule
Model & Log-likelihood & \# Params & AIC & BIC & p-value LRT \\
\midrule
G0 & -38,290 & 67 & 76,713 & 77,174 & $-$ \\ 
G1 & -38,279 & 68 & 76,695 & 77,162 & $7.0\times10^{-6}$ \\ 
G2 & -38,248 & 69 & 76,634 & 77,108 & $2.1\times10^{-15}$ \\ 
G3 & -38,221 & 70 & 76,582 & 77,063 & $2.2\times10^{-13}$ \\ 
G4 & -38,191 & 71 & 76,523 & 77,011 & $6.1\times10^{-15}$ \\ 
\bottomrule
\end{tabular}
\caption{Goodness of fit measures for the goal models (the last column is the p-value of the likelihood ratio test comparing with the previous model)}
\label{tab:aic_goals}
\end{table}

Tables~\ref{tab:parameters_goal_model} and \ref{tab:alpha_beta} display the exponentials of the estimated parameters $\xi$ for goal intensities. The exponentials act by multiplying the intensities so that values larger than $1$ increase the intensities while values in the range $(0,1)$ reduce them. We end this section by interpreting the parameter estimates for the outperforming model G4.

The home advantage parameter of $1.5140$ in model G4 means that the home team has a goal intensity 51.40\% higher than the visiting team, assuming all other parameters are equal. In other words, if one could replicate the game's circumstances, the home team would score, on average, 51.40\% more goals than the away team. Similarly, the half time estimate increases the goal rate on the second half by 20.62\% for both teams.

To interpret estimate of the coefficient of the value regressor, consider a situation where the two teams have exactly the same parameters playing against each other with no home advantage. For simplicity, we also assume the main squad of the home team is twice as valuable as the away team's squad. In this situation, the home team's goal rate is multiplied by $1.1454^{\log{(2)}} = 1.0987$, while the away team's goal rate is multiplied by $1/1.0987 = 0.9102$. Likewise, if the home team's squad is 10 times more valuable than the away team's, the former's goal rate is multiplied by 1.3670.

Under model G4, we find that the team that is leading by one goal has its goal rate multiplied by 0.9082 and when it is leading by two goals, its rate is multiplied by $(0.9082)^2 = 0.8248$. On the other hand, a team that is losing the match by one goal has its goal rate multiplied by $1/0.9082 = 1.1011$ or by $1/(0.9082)^2 = 1.2124$ if its two goals behind. Recall that an estimate smaller than 1 reduces the goal intensity. Thus, the goal intensity tends to decrease for the leading team and increase for the losing team.

When a team receives a red card, the estimates of model G4 imply the other team's goal intensity is multiplied by 1.4385 while the team that got the red card has its intensity decreased by 30.48\%.

\begin{table}[ht]
\label{modelos_gols}
\centering
\begin{tabular}{cccccc}
\toprule
Model & Home & Values & Half time & Goals & Red cards \\
\midrule
G0 & 1.4764 &  &  &  &  \\ 
G1 & 1.4661 & 1.1267 &  &  &  \\ 
G2 & 1.4662 & 1.1267 & 1.2078 &  &  \\ 
G3 & 1.5229 & 1.1407 & 1.2136 & 0.9169 &  \\ 
G4 & 1.5140 & 1.1454 & 1.2062 & 0.9082 & 1.4385 \\ 
\bottomrule
\end{tabular}
\caption{Parameter estimates for the goal models. The values presented are the exponentials of the coefficients $\xi$.}
\label{tab:parameters_goal_model}
\end{table}

\begin{figure} [ht] 
\caption{Ratio of attack and defence parameters vs average placement in the league}
\label{fig:attack_defence_position} 
\centering
\includegraphics[width=0.95\textwidth]{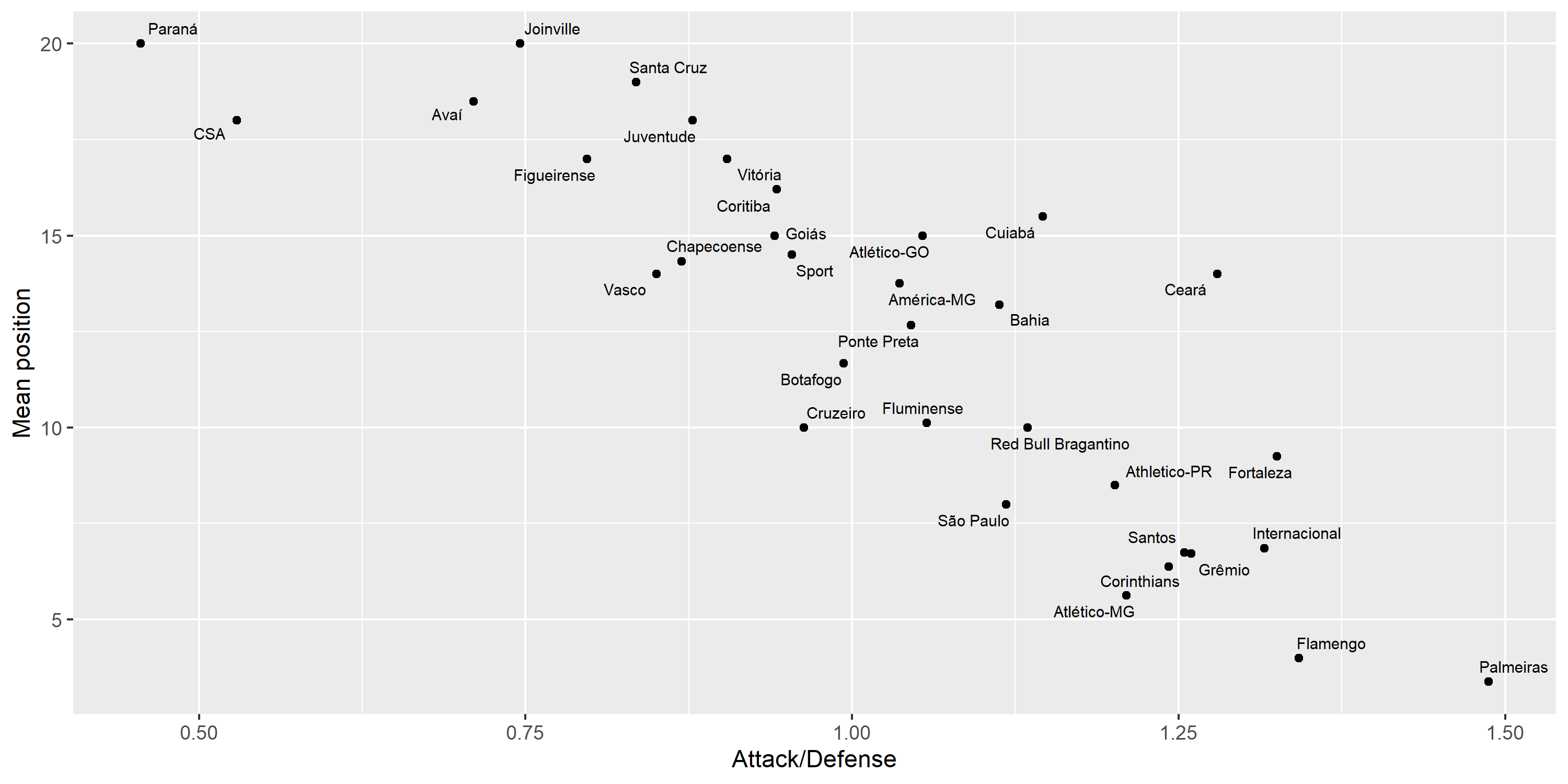}
\end{figure}

Regarding the team specific attack and defence parameters, one should first notice that a team's goal rate is proportional to both its own attack parameter and its opponent's defence parameter. Therefore one possible measure of a team's strength is the ratio between its attack and defence parameters. Table \ref{tab:alpha_beta} presents these parameters for all teams in our sample. Palmeiras is the team with highest ratio of attack and defence, followed by Flamengo, which are the two teams with the highest mean market value and also the best placement in the league; see Table \ref{tab:value}. Figure~\ref{fig:attack_defence_position} provides a graphical illustration of the relationship. Assuming the model has good predictive power (see Section~\ref{sec:forecast}), a team with a higher attack/defence ratio is expected to win more games, gather more points and place higher on the league.

\begin{table}[ht]
\centering
\begin{tabular}{rlccc}
  \toprule
 & Team & Attack & Defense & Attack/Defence \\ 
  \midrule
1 & Palmeiras & 0.1199 & 0.0806 & 1.4872 \\ 
2 &   Flamengo & 0.1187 & 0.0885 & 1.3420 \\ 
3 &   Fortaleza & 0.1043 & 0.0787 & 1.3254 \\ 
4 &   Internacional & 0.0986 & 0.0749 & 1.3157 \\ 
5 &   Ceará & 0.0934 & 0.0730 & 1.2797 \\ 
6 &   Grêmio & 0.1035 & 0.0821 & 1.2597 \\ 
7 &   Santos & 0.1028 & 0.0819 & 1.2545 \\ 
8 &   Corinthians & 0.0955 & 0.0769 & 1.2424 \\ 
9 &   Atlético-MG & 0.1161 & 0.0959 & 1.2101 \\ 
10 &   Athletico-PR & 0.0967 & 0.0805 & 1.2012 \\ 
11 &   Cuiabá & 0.0776 & 0.0677 & 1.1462 \\ 
12 &   Red Bull Bragantino & 0.1105 & 0.0974 & 1.1343 \\ 
13 &   São Paulo & 0.0950 & 0.0850 & 1.1180 \\ 
14 &   Bahia & 0.1023 & 0.0919 & 1.1126 \\ 
15 &   Fluminense & 0.0964 & 0.0912 & 1.0573 \\ 
16 &   Atlético-GO & 0.0903 & 0.0857 & 1.0539 \\ 
17 &   Ponte Preta & 0.0965 & 0.0923 & 1.0453 \\ 
18 &   América-MG & 0.0826 & 0.0797 & 1.0365 \\ 
19 &   Botafogo & 0.0867 & 0.0872 & 0.9937 \\ 
20 &   Cruzeiro & 0.0828 & 0.0860 & 0.9632 \\ 
21 &   Sport & 0.0901 & 0.0945 & 0.9541 \\ 
22 &   Coritiba & 0.0864 & 0.0917 & 0.9424 \\ 
23 &   Goiás & 0.0977 & 0.1039 & 0.9405 \\ 
24 &   Vitória & 0.1029 & 0.1138 & 0.9044 \\ 
25 &   Juventude & 0.0843 & 0.0960 & 0.8779 \\ 
26 &   Chapecoense & 0.0862 & 0.0991 & 0.8695 \\ 
27 &   Vasco & 0.0826 & 0.0971 & 0.8504 \\ 
28 &   Santa Cruz & 0.1085 & 0.1300 & 0.8347 \\ 
29 &   Figueirense & 0.0759 & 0.0952 & 0.7970 \\ 
30 &   Joinville & 0.0644 & 0.0864 & 0.7457 \\ 
31 &   Avaí & 0.0728 & 0.1025 & 0.7101 \\ 
32 &   CSA & 0.0563 & 0.1064 & 0.5288 \\ 
33 &   Paraná & 0.0454 & 0.0998 & 0.4552 \\ 
   \bottomrule
\end{tabular}
\caption{Attack and defense parameters for model G4. The values presented are the exponentials of the coefficients $\xi$.}
\label{tab:alpha_beta}
\end{table}

\subsection{Stoppage time models}

We study six different stoppage time models, named S0--S5. The parameter estimates are presented in Table~\ref{tab:stoppage_time_regressors}.

\begin{table}[h]
    \centering
    \begin{tabular}{ccccccc|c}
    \toprule
         Model & \text{Constant} & \text{Goals} & \text{Red cards} & \text{Goal difference} & \text{Restrictions}\\
    \midrule
    S0 & \ding{51} &  &   &  &  \\
    S1 & \ding{51} & \ding{51} &   &  & \ding{51}\\
    S2 & \ding{51} & \ding{51} &   &  & \\
    S3 & \ding{51} &  & \ding{51}   &  & \ding{51}\\
    S4 & \ding{51} &  & \ding{51} &  & \\
    S5 & \ding{51} &  & \ding{51} & \ding{51} & \\
    \bottomrule
    \end{tabular}
    \caption{Stoppage time regressors. Model S1 has the restriction that the goals parameter is the same on the first and second halves. Model S3 has a similar restriction, where the coefficient for red cards on the two halves is the same.}
    \label{tab:stoppage_time_regressors}
\end{table}

From model S0's estimates, on average, the first half receives 2.6680 minutes of extra play time, while the second half is usually given 4.8066 extra minutes. As discussed in Section \ref{ssec:regressors_stoppage_time}, we also test the impact of including the total number of goals as a regressor for the stoppage time. 

Model S1 adds this regressor, with the restriction that its coefficients for both halves must be the same. On Model S1, the constants estimated in Model S0 are mostly unaffected and the goal's coefficient is slightly lower than one, meaning that each goal scored decreases the expected extra time by 1.25\%. Model S2, removes the restriction that the coefficients of the goals' regressors need to be the same on both halves. Nonetheless, its estimates are very close and still around 1, meaning that the total number of goals scored have no effect on the stoppage time given by the referee.

Similarly to models S1 and S2, models S3 and S4 differ only on the fact that model S4 allows for different coefficients for the total number of red cards parameter on the two halves of the matches. We interpret the effect of these estimates on a synthetic example. Let us suppose during the regular 45 minutes of the second half there are 2 red cards awarded to the home team and one to the away team (red cards awarded on the first half are ignored when computing the regressor for the second half). Then, on average, Model S3 expects $4.7367 \times 1.1104^{2+1} = 6.4851$ minutes of added time. Similarly, under Model S4, the best estimate is to expect $4.7688 \times 1.0597^{2+1} = 5.6749$ minutes to be added to the match.

In Model S5 we keep all regressors from Model S4 and add the one that measures if the match result is close (difference of one goal or less). Adding the assumption of a tied match to the example of the previous paragraph, Model S5 expects an extra time of $3.9640 \times 1.0506^{2+1} \times 1.2814 = 5.8902$ minutes. If, instead, the match is $3 \times 0$, this model expects 1.2935 minute less for extra time than if the match was tied. 

\begin{table}[ht]
\centering
\begin{tabular}{cccccccc}
\toprule
{\scriptsize Model} & {\scriptsize Const 1} & {\scriptsize Const 2} & {\scriptsize Goals 1} & {\scriptsize Goals 2} & {\scriptsize Red cards 1} & {\scriptsize Red cards 2} & {\scriptsize Goal diff 2} \\
\midrule
S0 & 2.6680 & 4.8066 &  &  &  &  &  \\ 
S1 & 2.6992 & 4.8743 & 0.9875 & 0.9875 &  &  &  \\ 
S2 & 2.5629 & 5.0153 & 1.0432 & 0.9621 &  &  &  \\ 
S3 & 2.6574 & 4.7367 &  &  & 1.1104 & 1.1104 &  \\ 
S4 & 2.6233 & 4.7688 &  &  & 1.4694 & 1.0597 &  \\ 
S5 & 2.6234 & 3.9640 &  &  & 1.4693 & 1.0506 & 1.2814 \\ 
\bottomrule
\end{tabular}
\caption{Parameter estimates for the stoppage time models.}
\label{tab:stoppage_time_parameters}
\end{table}


Differently from the goals model, the proposed stoppage time models are not all nested. However, when grouping together models S0$-$S1$-$S2 and S0$-$S3$-$S4$-$S5, we find two collections of nested models. Within the first collection, all goodness of fit measures computed in Table \ref{tab:stoppage_time_gof} (log-likelihood, AIC and BIC) suggest that Model S2 has the best fit. On the second collection of models, S0$-$S3$-$S4$-$S5, we have that Model S5 provides the best fit to the observed data.  As these measures are not perfectly suitable for comparison of non-nested models, one can not say Model S5 is better than Model S2, but for the remainder of the paper we present results based on Model S5 only.

\begin{table}[ht]
\centering
\begin{tabular}{ccccc}
\toprule
Model & Log-likelihood & \# Params & AIC & BIC \\ 
\midrule
S0 & -11,514 & 2 & 23,032 & 23,045 \\ 
S1 & -11,512 & 3 & 23,030 & 23,050 \\ 
S2 & -11,496 & 4 & 22,999 & 23,026 \\ 
S3 & -11,500 & 3 & 23,005 & 23,026 \\ 
S4 & -11,481 & 4 & 22,971 & 22,998 \\ 
S5 & -11,399 & 5 & 22,808 & 22,842 \\ 
\bottomrule
\end{tabular}
\caption{Information criteria for the stoppage time models. Note that models S0$-$S1$-$S2 and S0$-$S3$-$S4$-$S5 are nested, but the collection S0$-$S5 is not nested.}
\label{tab:stoppage_time_gof}
\end{table}


\subsection{Red cards model}

We next estimate the two parameters for the red card model from Section~\ref{sec:red-cards}. The estimated values are presented in Table~\ref{tab:red_cards_estimates}. At minute 15, the average intensity of red cards for the home team is $e^{-12.6054} \times 15^{1.4275} = 1.6010 \times 10^{-4}$ red cards per minute. Likewise, at minute 75, the intensity is estimated as $1.5929\times 10^{-3}$ for the home team, which is about 10 times higher than the intensity at minute 15. Although similar, the away rate is lower than the home red cards rate before minute 8 and after that, it is consistently higher. For example, at minute 75, the away red cards rate is $e^{-13.0132} \times 75^{1.6272} = 2.5092 \times 10^{-3}$, which is about 1.5 times higher than the home team.

\begin{table}[ht]
\centering
\begin{tabular}{lr}
\toprule
Regressor & Estimate \\ 
\midrule
Home constant & -12.6054  \\ 
Home time & 1.4275  \\ 
Away constant & -13.0132  \\ 
Away time & 1.6272 \\ 
\bottomrule
\end{tabular}
\caption{Parameter estimates for the red card model.}
\label{tab:red_cards_estimates}
\end{table}

\section{Forecasting} \label{sec:forecast}

In order to evaluate the forecasting ability of the goals models for a specific match, we use all information available at the day prior to the match to estimate the model parameters. During the match, we produce on-line estimates, which are updated as time evolves and events (red cards or goals) happen. 

When forecasting a specific match, we simulate the four processes (goals and red cards for both teams), select the one that will happen first, update the state of the match (current score and time) and repeat. For each half, we do that until minute 45, simulate the stoppage time and repeat that algorithm in the stoppage time. For the red cards processes that are nonhomogeneous, the inversion method described in \cite{cinlar} is used.

To have enough information to estimate the models, we exclude from the rolling forecasting window, the first four matches of each team and also the season 2015.

We test five goals models (G0$-$G4) combined with the red cards model (R) and the chosen stoppage time model (S5). The data set used for this forecasting exercise contains a total of 2\,611 matches, down from the initial 3\,039. For each match, we simulate 100\,000 scenarios starting at minutes 0, 15, 30, 45, 60 and 75. 

When simulating the match events after time 0, we use all available information from that match up to the minute when the simulation is being performed. For example, suppose the only events in a match are a goal scored by the home team at minute 31 and a red card awarded to the away team at minute 61. If forecasts are computed at minute 45, only information on the home goal will be used. On the other hand, if forecasts are computed at minute 75, both the home goal and the away red card will be used by the model.

\subsection{Likelihood of the result and exact score data}
In order to evaluate the fit of the models to the observed data, we define as ``result" the outcome of the match as Home Win $-$ Draw $-$ Away Win. The ``exact score" is the number of goals scored by each team, for example 2 (Home Team) $\times$ 0 (Away Team).

Knowing the outcome of the match (either the ``result" or the ``exact score"), we compute the likelihood of the observations using the conditional probability at time $t$ according to model $m$. This probability is denoted by $p_m^k(t)$. If a match ends with a Home Win, the likelihood of the ``results" is computed interpreting $p_m^k(t)$ as the probability, at time $t$, of observing a Home Win at the end of the match. Note that the probabilities $p_m^k(t)$ can be computed numerically by simulation.

The likelihood according to model $m$ of observing the outcomes of all the $K$ matches is then given by
\[
\L_m(t):= \prod_{k=1}^K p^k_m(t).
\]

\begin{figure} [ht] 
\caption{Differences of (a) the models' log-likelihoods for ``results" at minutes 0, 15, 30, 45, 60 and 75 and (b) the log-likelihood of Model G0S5R, both normalized by 2\,611, the number of matches that were simulated.}
\label{fig:likelihood_results} 
\centering
\includegraphics[width=0.95\textwidth]{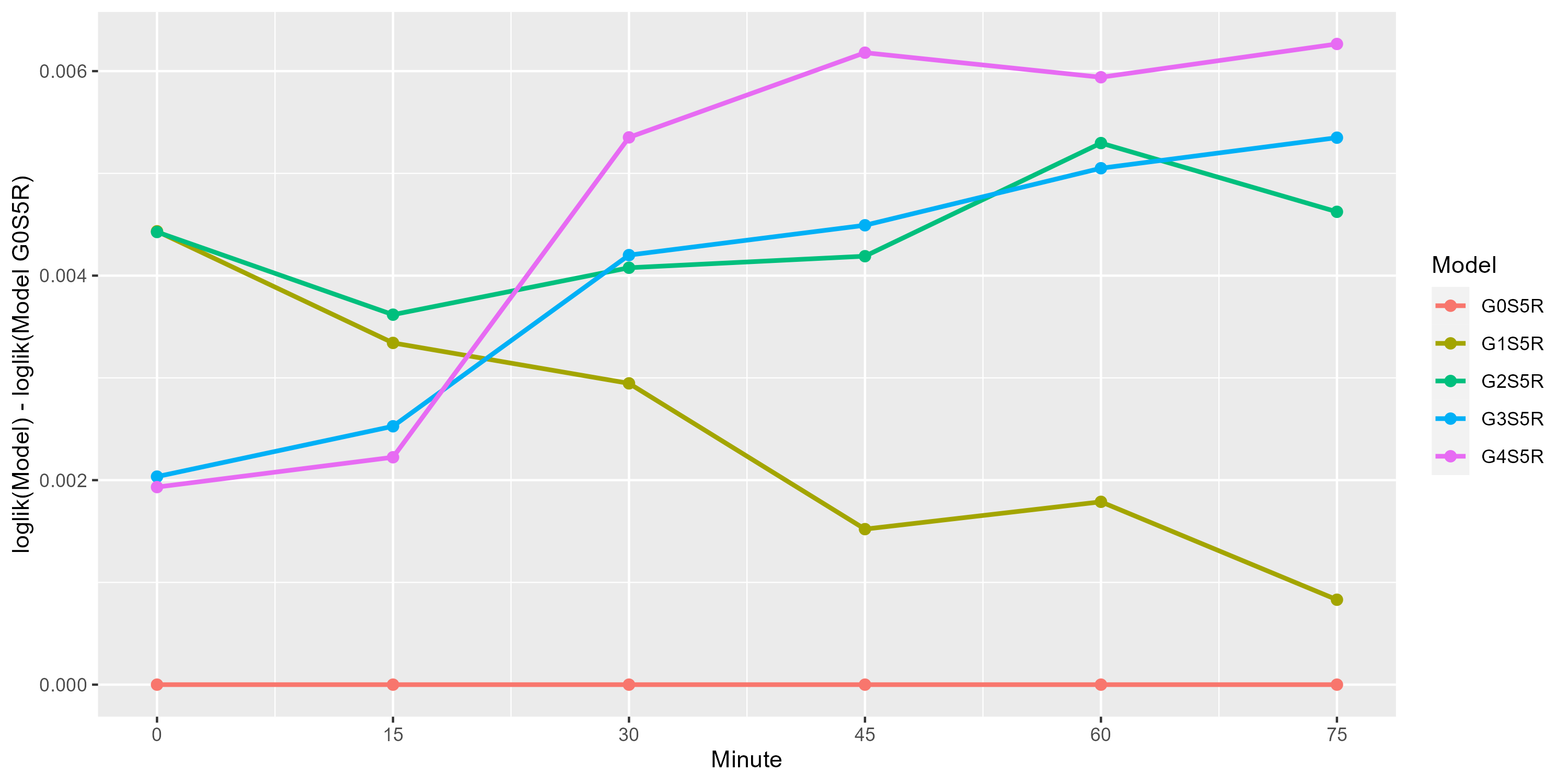}
\end{figure}

\begin{figure} [ht] 
\caption{Differences of (a) the models' log-likelihoods for ``exact scores" at minutes 0, 15, 30, 45, 60 and 75 and (b) the log-likelihood of Model G0S5R, both normalized by 2\,611, the number of matches that were simulated.}
\label{fig:likelihood_scores} 
\centering
\includegraphics[width=0.95\textwidth]{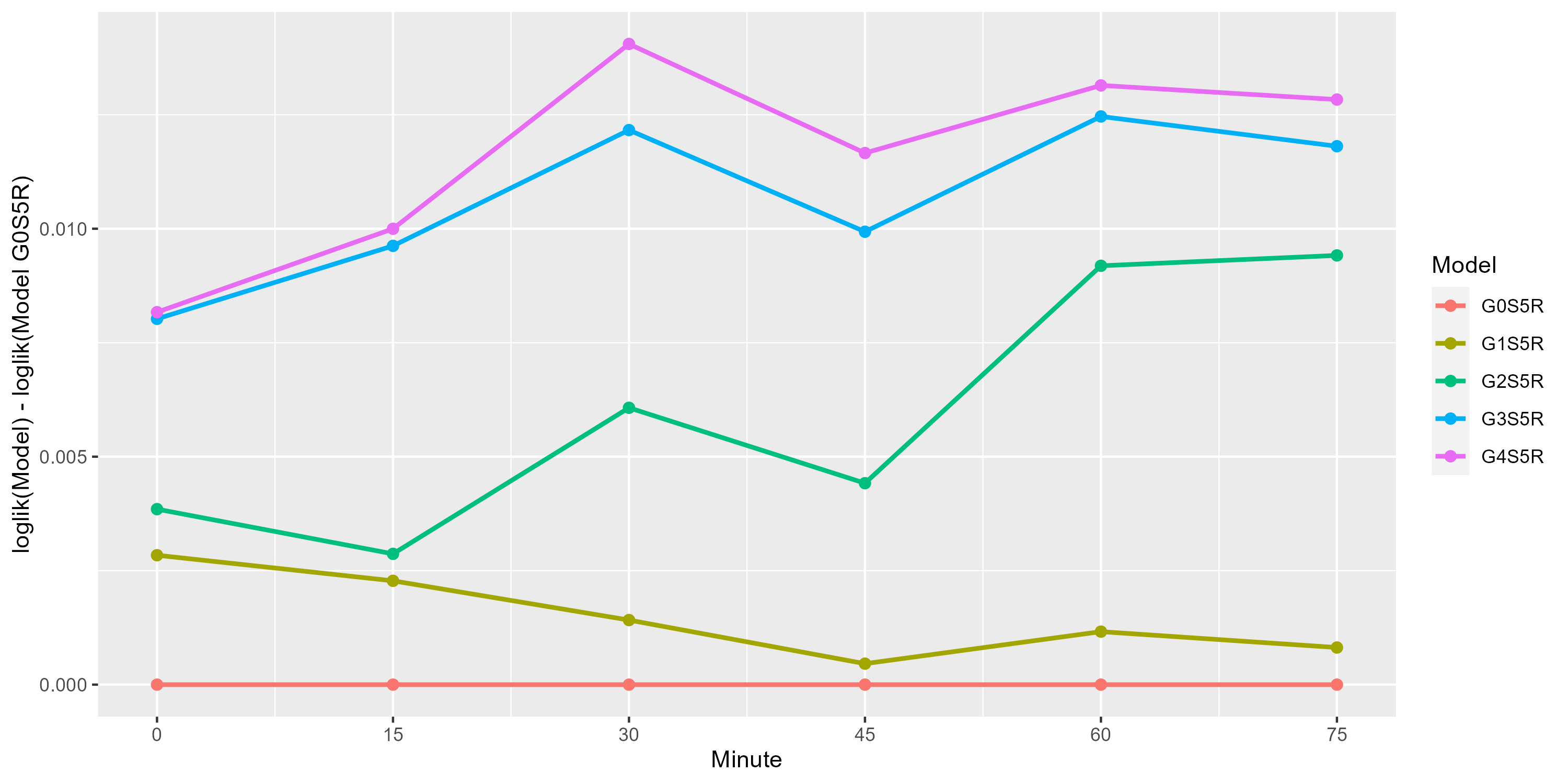}
\end{figure}

Figures \ref{fig:likelihood_results} and \ref{fig:likelihood_scores} illustrate the differences of (a) the models' log-likelihoods at minutes 0, 15, 30, 45, 60 and 75 and (b) the log-likelihood of Model G0S5R, both normalized by 2\,611, the number of matches that were simulated.

The first thing one should notice from Figure \ref{fig:likelihood_results} is the fact that the simplest model tested, G0S5R, is worse than all others regardless of the moment of the match the prediction is being performed. When adding value of the teams to G0S5R, leading to model G1S5R, we observe a substantial improvement on the predictive power of this model at the beginning of the matches. Nonetheless, the performance of model G1S5R diminishes over time, to the point that at minute 75 its likelihood is almost the same as model G0S5R. On possible explanation is that the covariate measuring the value of the roster is computed for the initial lineup. As the match evolves and substitutions naturally occur, this variable tends to become obsolete and not to bring any predictive power. The remaining models in Figure \ref{fig:likelihood_results} (G2S5R, G3S5R and G4S5R) include a half time variable and show remarkable improvements in the likelihoods, especially after minute 45. In particular, model G4S5R, which also includes the goals variable and the red cards variable, has the best likelihood to predict ``results" if predictions are being performed from minute 30 onwards.

Similarly to Figure \ref{fig:likelihood_results}, Figure \ref{fig:likelihood_scores} presents the difference in likelihoods, now computed for ``exact scores". In this case, regardless of the minute in which the predictions are being computed, the most complex models always perform better than the simplest ones.

\subsection{Empirical probabilities of events}

Let us denote by $E$ an event of interest, such as ``only one or two goals are scored by the home team" or ``0 or 4 goals are scored by the away team". We will denote by $P_m^k(E)$ the probability according to model $m$ of event $E$ in game $k$. The expected number of games $N_E$ in which event $E$ occurs is then given by
\[
E[N_E] = \sum_{k=1}^K P_m^k(E).
\]
We can compare these numbers with the observed number of games where $E$ happens.

A challenging and recurring theme when estimating exact scores is the fact that Draws are usually underestimated. One notable example is the seminal paper \cite{maher1982modelling}. In that paper, two models are proposed for home and away goals (stoppage time and red cards are ignored): the first one is similar to G0 with $k$ playing the role of the home advantage parameter; the second one is similar to the first but assuming a bivariate Poisson distribution for home and away goals. We remind the reader that, as discussed in Section \ref{ssec:dgoalmodels}, in models G3 and G4 the home and away goals are not independent.

The data used in \cite{maher1982modelling} consists of 462 matches, with a total of 129 Draws (27.92\%). From Table 6 in \cite{maher1982modelling}, the simpler independent Poisson model estimates a total of 113 Draws (24.46\%) while their proposed bivariate Poisson model predicts 126.2 (27.32\%) matches would end up as Draws. 

The first line of Table \ref{tbl:gof_results} presents the observed proportion of results (Home Win, Draw and Away Win) across all 2\,611 matches we performed simulations for. In our data set, 27.42\% of the matches end as draws. The following lines show the difference between these observed proportions to the proportions simulated by each one of the models before the beginning of the matches (at minute 0). The average number of Draws under Model G4S5R is $0.2748 = 0.2742 + 0.0006$, and the difference to the observed proportion of Draws is significantly smaller then the one observed by \cite{maher1982modelling}. In particular, one should notice that models with more covariates tend to improve the estimated number of Draws.


On top of estimating the proportion of Home/Away wins and Draws, we can also compare the proportion of matches that ended up with a certain number of goals scored by the home and away teams. These information is described in Tables \ref{tbl:gof_home_score} and \ref{tbl:gof_away_score}, respectively, in a similar fashion to Table \ref{tbl:gof_results}.

As expected, the more covariates are included in the goals models, the better the models become. Models G1S5R and G2S5R do not differ significantly when predicting the final number of goals scored by the home/away teams. This is expected, as we start the simulation at time $0$ and run it until the end of the match. This would not be the case if we were analysing the estimates during the match. When analysing the full match, model G2S5R behaves as if part of the G1S5R's goal intensity from the first half is moved to the second. Models G3S5R and G4S5R present similarly good results.

\begin{table}[ht]
\caption{Results probabilities: observed and estimated.}
\centering
\begin{tabular}{lrrr}
\toprule
& Home & Draw & Away \\ 
\midrule
Observed & 0.4780 & 0.2742 & 0.2478 \\ 
G0S5R & +0.0048 & $-$0.0114 & +0.0066 \\ 
G1S5R & +0.0052 & $-$0.0133 & +0.0081 \\ 
G2S5R & +0.0052 & $-$0.0134 & +0.0082 \\ 
G3S5R & +0.0072 & +0.0003 & $-$0.0075 \\ 
G4S5R & +0.0076 & +0.0006 & $-$0.0082 \\ 
\bottomrule
\end{tabular}
\label{tbl:gof_results}
\end{table}

\begin{table}[ht]
\caption{Probabilities of the number of goals: home team.}
\centering
\begin{tabular}{lrrrrrr}
\toprule
& 0 & 1 & 2 & 3 & 4 & 5 \\ 
\midrule
Observed & 0.2332 & 0.3658 & 0.2413 & 0.1122 & 0.0368 & 0.0107 \\ 
G0S5R & +0.0236 & $-$0.0336 & $-$0.0086 & +0.0020 & +0.0073 & +0.0092 \\ 
G1S5R & +0.0244 & $-$0.0351 & $-$0.0097 & +0.0022 & +0.0081 & +0.0101 \\ 
G2S5R & +0.0244 & $-$0.0351 & $-$0.0096 & +0.0022 & +0.0080 & +0.0101 \\ 
G3S5R & +0.0071 & $-$0.0252 & +0.0046 & +0.0049 & +0.0045 & +0.0040 \\ 
G4S5R & +0.0060 & $-$0.0251 & +0.0054 & +0.0053 & +0.0045 & +0.0040 \\ 
\bottomrule
\end{tabular}
\label{tbl:gof_home_score}
\end{table}

\begin{table}[ht]
\caption{Probabilities of the number of goals: away team.}
\centering
\begin{tabular}{lrrrrrr}
\toprule
& 0 & 1 & 2 & 3 & 4 & 5 \\ 
\midrule
Observed & 0.3822 & 0.3696 & 0.1838 & 0.0467 & 0.0142 & 0.0034 \\ 
G0S5R & +0.0252 & $-$0.0154 & $-$0.0180 & +0.0079 & +0.0000 & +0.0003 \\ 
G1S5R & +0.0258 & $-$0.0168 & $-$0.0182 & +0.0084 & +0.0003 & +0.0006 \\ 
G2S5R & +0.0257 & $-$0.0168 & $-$0.0182 & +0.0083 & +0.0003 & +0.0006 \\ 
G3S5R & +0.0126 & $-$0.0040 & $-$0.0123 & +0.0065 & $-$0.0019 & $-$0.0009 \\ 
G4S5R & +0.0117 & $-$0.0034 & $-$0.0120 & +0.0066 & $-$0.0020 & $-$0.0009 \\ 
\bottomrule
\end{tabular}
\label{tbl:gof_away_score}
\end{table}

\subsection{Minute-by-minute prediction}
Taking the likelihood as the measure of goodness-of-fit, we now present the minute-by-minute predictive results for two different matches, computed with 100\,000 simulations.

\begin{figure} [ht] 
\caption{Minute-by-minute forecasts (under model G4S5R) of exacts scores for Ceará x Paraná (2018) }
\label{fig:ceara_parana} 
\centering
\includegraphics[width=0.95\textwidth]{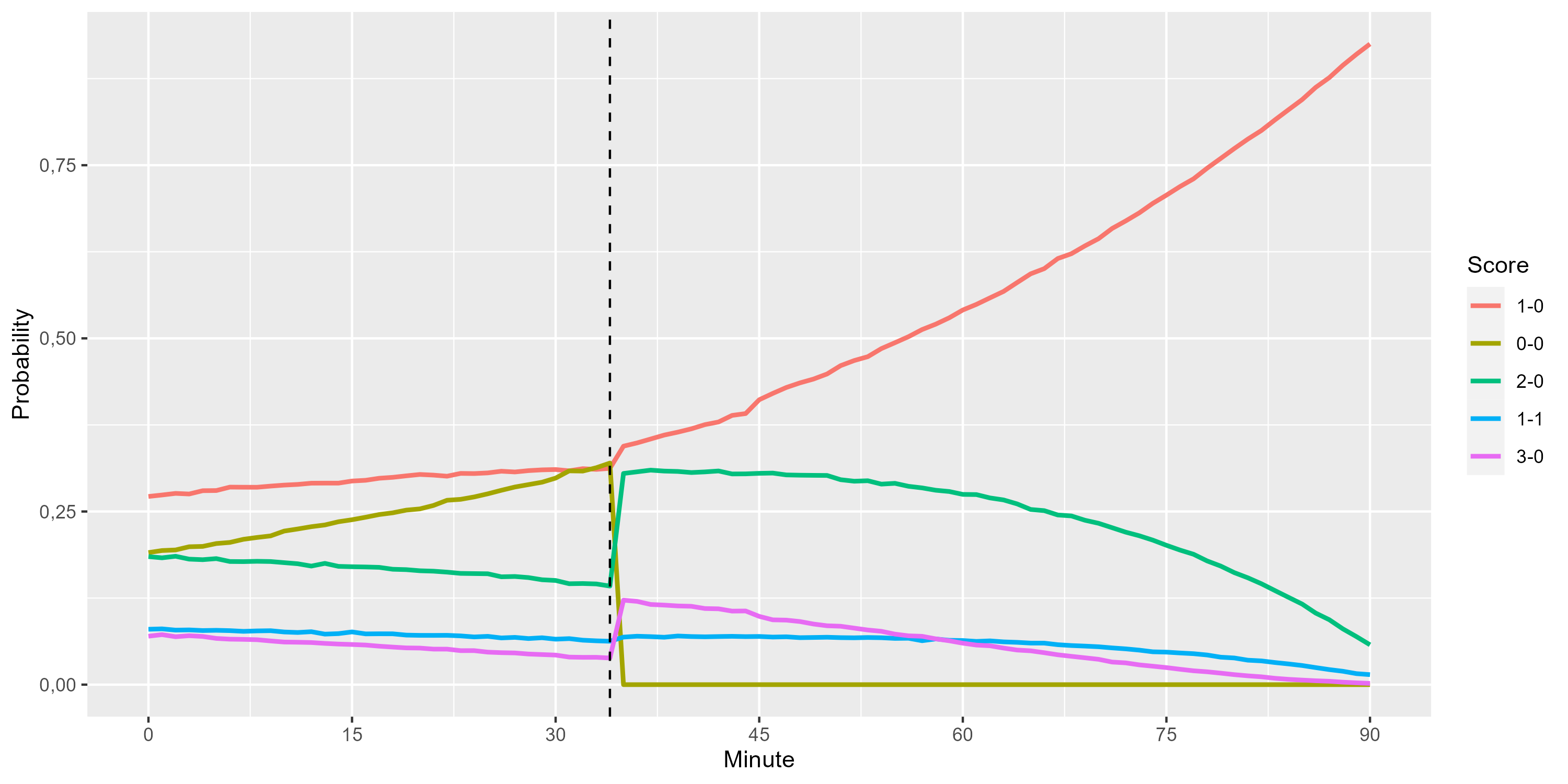}
\end{figure}

First, we analyse forecasts made by model G4S5R at time 0 for the match Ceará x Paraná of the 2018 league. This particular match, which ended as 1-0, has the highest likelihood (at time 0) for the exact score observed. Figure \ref{fig:ceara_parana} presents the probabilities for the 5 most likely exact scores at minute 0 along the match. For example, taking into account all the information available until minute 15, the model predicts that 1-0 is a way more likely end game score than 3-0. The vertical dashed line (at minute 34) represents the moment when the home team scores its first (and only) goal. Before that, the probability of the match ending as a 0-0 draw increases rapidly, but goes to zero as soon as the goal is scored. On the other hand, scores like 2-0 and 3-0 becomes more likely. As no other goals are scored, as time goes by the 1-0 probability goes up fast, while all other scores' probabilities move towards to zero. It should be noticed that even though the model takes into account the added stoppage time, Figure \ref{fig:ceara_parana} only shows forecasts up to minute 90, hence the most likely score doesn't have probability exactly equal to 1 at the end of the plot (which is not necessarily the end of the match).

\begin{figure} [ht] 
\caption{Minute-by-minute forecasts (under model G4S5R) of end game results for Ponte Preta x Vitória (2017)}
\label{fig:ponte_preta_vitoria} 
\centering
\includegraphics[width=0.95\textwidth]{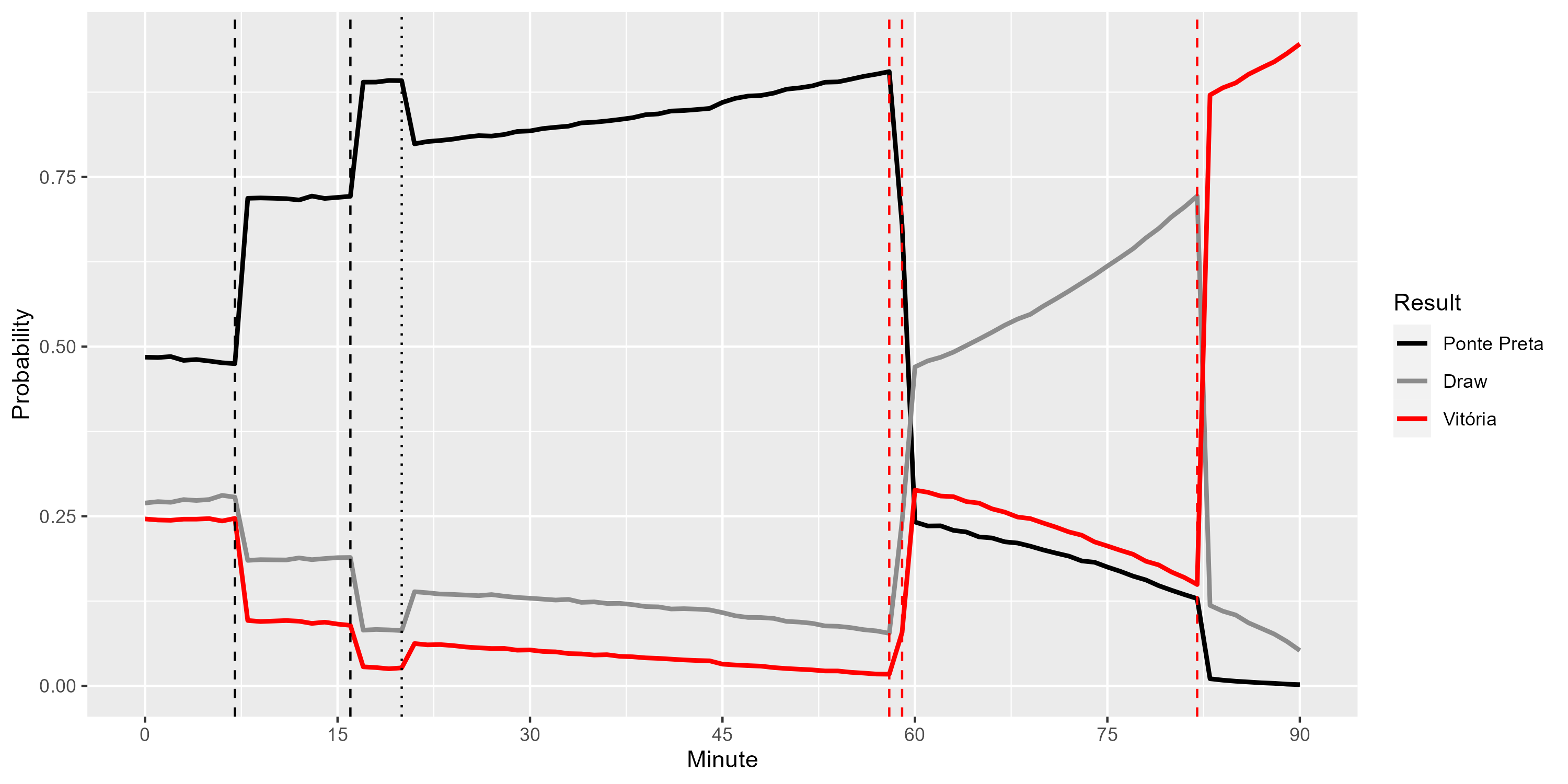}
\end{figure}

Next we analyse the effect of adding red cards as a covariate in the model. To do so, we select the match with the highest proportional difference of the likelihoods for the end result (Home/Away wind, Draw) of models G3S5R and G4S5R at minute 45. In other words, we select the match in which, at minute 45, the impact of adding red cards information into the modelling is the greatest. The match in question is Ponte Preta x Vitória in the 2017 league and Figure \ref{fig:ponte_preta_vitoria} presents the forecasts, minute-by-minute, of model G4S5R for the end result of the match. Dashed vertical lines denote goals scored (black for the home team, red for the away), while dotted lines represent red cards. This match had the following events: Ponte Preta goals (7' and 16'); Ponte Preta red card (20'); Vitória goals (58', 59' and 82'). 

At minute 0, the probability of a home win is almost twice the probability of an away win, while a draw is given 27\% chance. The home win probability increases with each goal scored and decreases slightly after the red card. Then, as nothing else happens in the match, the home win probability steadily grows towards one until it falls abruptly after the two goals scored by the away team. After these two goals, the most likely result is a draw until the moment when the away team scores a late goal, making the away team win probability increase from 15\% (at 81') to 87\% (at 83').

\bibliographystyle{plain}
\bibliography{refs}

\end{document}